\documentclass[useAMS,usenatbib]{mn2e}

\usepackage{graphicx}
\usepackage{graphicx}
\usepackage{color}
\usepackage{journal_names}
\usepackage{subfig}
\usepackage{float}

\title[OmegaWhite Survey]
  {The OmegaWhite Survey for Short-Period Variable Stars I: Overview and First Results}
\author[S. Macfarlane et al.]
  {S.A~Macfarlane$^{1,2}$\thanks{Email: S.Macfarlane@astro.ru.nl},
  R.~Toma$^3$,
  G.~Ramsay$^3$,
  P.J~Groot$^1$,
  P.A~Woudt$^2$,
  J.E~Drew$^4$, 
  \newauthor
  G.~Barentsen$^4$,
  J.~Eisl\"offel$^5$,  
  \\
  $^1$Department of Astrophysics/IMAPP, 
      Radboud University, 
      P.O. Box 9010,
      6500 GL Nijmegen,
      The Netherlands \\
  $^2$Astrophysics,Cosmology and Gravity Centre, 
      Department of Astronomy, 
      University of Cape Town, 
      Private Bag X3, \\
      Rondebosch 7701, 
      South Africa \\      
  $^3$Armagh Observatory, 
      College Hill, 
      Armagh, 
      BT61 9DG,
      Northern Ireland\\
  $^4$School of Physics, Astronomy \& Mathematics,
      University of Hertfordshire,
      College Lane,
      Hatfield, 
      Hertfordshire,
      AL10 9AB, 
      U.K.\\
  $^5$Th\"uringer Landessternwarte, 
      Sternwarte 5,
      D-07778 Tautenburg, 
      Germany \\
 } 

\pagerange{\pageref{firstpage}--\pageref{lastpage}} \pubyear{2015}

\def\LaTeX{L\kern-.36em\raise.3ex\hbox{a}\kern-.15em
    T\kern-.1667em\lower.7ex\hbox{E}\kern-.125emX}

\begin{document}

\label{firstpage}

\maketitle

\begin{abstract}

We present the goals, strategy and first results of the OmegaWhite survey: a wide-field high-cadence $g$-band synoptic survey which aims to unveil the Galactic population of short-period variable stars (with
periods $<$ 80 min), including ultracompact binary star systems and stellar pulsators. The ultimate goal of OmegaWhite is to cover 400 square degrees along the Galactic Plane reaching a depth of $g = $ 21.5 mag (10$\sigma$),
using OmegaCam on the VLT Survey Telescope (VST). The fields are selected to overlap with surveys such as the Galactic Bulge Survey (GBS) and the VST Photometric H$\alpha$ Survey of the Southern Galactic Plane (VPHAS+) for multi-band colour information. Each field is observed using 38 exposures of 39 s each, with a median cadence of $\sim$2.7 min for a total duration of two hours. Within an initial 26 square degrees, we have extracted the light curves of 1.6 million stars, and have identified 613 variable candidates which satisfy our selection criteria. Furthermore, we present the light curves and statistical properties of 20 sources which have the highest-likelihood of being variable stars. One of these candidates exhibits the colours and light curve properties typically associated with ultracompact AM\,CVn binaries, although its spectrum exhibits weak Balmer absorption lines and is thus not likely to be such a binary system. We also present follow-up spectroscopy of five other variable candidates, which identifies them as likely low-amplitude $\delta$ Sct pulsating stars.

\end{abstract}

\begin{keywords}
 surveys -- binaries: close -- Galaxy:bulge -- methods: observational -- methods: data analysis -- techniques: photometric.
\end{keywords}

\section{Introduction}

\setcounter{table}{0}
\begin{table*}
\centering
  \caption{Overview of recent high-cadence synoptic surveys}
  \begin{tabular}{lcccc}
   \hline
   Survey & Field Location & Total Sky Coverage & Depth & Variability sensitivity \\  
    & (deg) & (deg$^{2}$) & ($V$ mag) & (min) \\
   \hline
   OmegaWhite$^{a}$ & $|$b$|$ $\leq$ 10 & 400 & 21.5 & \textgreater\ 6 \\
   RATS$^{b}$ & $|$b$|$ $\leq$ 30 & 46 & 22.5 & \textgreater\ 5 \\
   RATS-Kepler$^{c}$  & 6 $\leq$ $|$b$|$  $\leq$ 21 & 49 & 22.5 & \textgreater\ 5 \\
   FSVS$^{d}$  & $|$b$|$ $\geq$ 20 & 23 & 24.0 & \textgreater\ 24  \\
   DLS$^{e}$ &  $|$b$|$  \textless\ 10 & 21 & 25.5 &  \textgreater\ 15  \\
   Kepler$^{f}$ &  15 $\leq$ $|$b$|$  $\leq$ 25 & 116 & 20.0 &  \textgreater\ 1 or \textgreater\ 30  \\
   SuperWASP$^{g}$ & all sky & all sky & 15.0 & \textgreater\ 10  \\
   \hline
   \multicolumn{5}{p{12cm}}{$^{a}$This paper, $^{b}$Rapid Temporal Survey \citep{Ramsay2005}, $^{c}$\citep{Ramsay2014}, $^{d}$Faint Sky Variability Survey \citep{Groot2003}, Deep Lens Survey $^{e}$\citep{Becker2004}, $^{f}$ \citep{Borucki2010}, $^{g}$Wide Angle Search for Planets \citep{Pollacco2006}}\\
  \end{tabular}
  \label{surveycomp}
\end{table*}

Time domain astrophysics has been transformed over the last
decade. Whereas in the past, photometric data of individual objects
were painstakingly obtained using high speed photometry or dedicated
long term projects, now a whole series of synoptic projects have been
developed to observe large areas of sky over
short time-scales giving photometric data on thousands or millions of
objects. The diversity of goals of these projects is considerable, ranging
from detecting transiting exo-planets
\citep[e.g. the main aim of the `Super-WASP' project,][]{Pollacco2006} to discovering supernova
outbursts \citep[e.g. supernova discoveries in the Palomar Transient Factory (PTF),][]{Law2009}.

In the field of Galactic binary research, one key goal has been to discover individual systems with astrophysically interesting properties. Of particular interest are ultracompact binaries (UCBs) which have an orbital period (P$_{orb}$) of $\leq$ 70 min, implying that the
secondary star cannot be a main-sequence star
\citep{Rappaport1982}. Furthermore, these hydrogen-deficient objects
are predicted to be the strongest known sources of gravitational wave
radiation (GWR) in the passband of the satellite observatory {\sl eLISA}
\citep{Amaro-Seoane2013,Roelofs2007}, and as such are important calibrators that provide verification of the existence and detectability of GWR. Moreover, the evolution of these binary systems is influenced by the emission of GWR in addition to the mass transfer phase. Therefore, the study of UCBs will also help to answer key questions of late-stage binary evolution. Earlier estimates of their intrinsic numbers
suggested a relatively high foreground contribution from UCBs to the
gravitational wave signal from merging supermassive black
holes within the {\sl eLISA} band. However, in a series of papers using SDSS and PTF data
\citep{Roelofs2009,Rau2010,Carter2013,Levitan2015}, it is now clear that the
predicted number density of AM\,CVn binaries (semi-detached and mass
transferring UCBs) in the Solar neighbourhood is 5 $\pm$ 3 $\times$ 10$^{-7}
$pc$^{-3}$, a factor 50 lower than previous estimates by
\citet{Nelemans2001b}. Other surveys which have recently had success in discovering new AM\,CVn systems include the Catalina Real-Time Transient Survey \citep[CRTS][]{Drake2009}, and the All Sky Automated Survey for Supernovae \citep[ASAS-SN, e.g.][]{Wagner2014}.

The emission line method for finding AM\,CVn systems, on which
the SDSS work is based, is most sensitive to systems with P$_{orb}
\geq$ 30 min. More recently, the PTF survey has been identifying outbursting
AM\,CVns which have orbital periods shorter than this, but still greater than 22 min \citep{Levitan2015}. From the total number of 43 known AM\,CVns, only 6 have periods shorter than 20 minutes. It is those systems with the shortest orbital period (5 min $<$ P$_{orb} < $ 20 min) which are predicted to be the strongest emitters of gravitational waves.  Determining their population size is important for both the development of {\sl eLISA}
and for deducing the relative importance of the three postulated formation channels of these binaries \citep[for a detailed review, see][]{Solheim2010}. A way to identify those AM\,CVn stars with the
shortest orbital periods is through their photometric behaviour as they show a periodic modulation on, or close to, the orbital period \citep[e.g. the first PTF AM\,CVn discovered,][]{Levitan2011}. Variations are expected due to eclipses, ellipsoidal variations, irradiation, superhumps, or anisotropic disc hotspot emission in the system. The detection of these especially short-period systems is the main motivation behind the OmegaWhite survey.

However, short time-scale photometric variations can also originate from
physical changes within the internal structure or atmosphere of the
source. Examples of such sources include flare stars or fast pulsating
stars such as $\delta$ Scuti stars \citep[$\delta$ Sct,][]{Breger2000}
or SX Phoenicis variables \citep[e.g.][]{Rodriguez2000}. At even
shorter periods, rapidly oscillating Ap or Am star systems (roAp,
roAm), and pulsating white dwarfs (ZZ Ceti variables) exhibit periodic variations on amplitudes ranging from a percent or less up to several tens of percent on timescales of a few to tens of minutes.

$\delta$ Sct stars have A--F spectral types and are known to have short pulsation periods \citep[typically 0.02 - 0.25 days,][]{Chang2013}, with absolute magnitudes 3 - 5 mag fainter than Cepheids. However, they are believed to be the second most common variable stars in the Galaxy \citep{Breger1979}, and their short period makes them relatively easy to detect in high-cadence surveys \citep[e.g.][]{Ramsay2005}.  $\delta$ Sct stars can be used as precise distance tracers if the fundamental radial pulsation mode can be identified
\citep[e.g.][]{Peterson1999, McNamara2007}. For high-amplitude
$\delta$ Sct stars (HADS), where the pulsation amplitude is greater than 0.3
mag, it is expected that the dominant period will be due to the
fundamental or first-overtone radial mode.  The shortest period
$\delta$ Sct stars in the Galactic field have a dominant pulsation
period of $\sim$26\,min  \citep{Kim2010}. 

One survey that set out to discover short-period systems was the
Rapid Temporal Survey \citep[RATS,][]{Ramsay2005,Barclay2011}. RATS
was carried out using the 2.5-m Isaac Newton Telescope (INT) on La Palma and the 2.2-m Max Planck Gesellschaft telescope (MPG/ESO) between 2003--2010, and covered a total of 46 square degrees favouring fields near the Galactic plane. The fact that no AM\,CVn was discovered was (in retrospect) not a surprise since
the latest space densities imply that only a few systems (at best) are
expected in a survey covering this area. However, the survey did discover many new and interesting short-period variables, including a possible hybrid sdBV pulsator exhibiting long-period g-modes
\citep{Ramsay2006} and a dwarf nova discovered through its high-amplitude quasi-periodic oscillations (QPOs) in quiescence \citep{Ramsay2009}.

\setcounter{table}{1}
\begin{table*}
\centering
 \caption{OmegaWhite observation log for observations made between December 2011 and April 2015}
  \begin{tabular}{cccccc}
   \hline
   ESO  Period & Observation Dates& Allocated Time& Observed Fields& RA (J2000) & DEC (J2000) \\
    &  & (hrs) & (hrs) & (hh:mm) & (\degr:\arcmin)  \\
   \hline
   88 & Dec. 2011 to Apr. 2012 & 32 & 28 & 07:35 $-$ 08:25 & --30:00 $-$ --26:00   \\
   90   & Nov. 2012 to Mar. 2013 & 64 & 8 & 07:05 $-$ 08:25  & --30:00 $-$ --25:00  \\
   91   & Apr. 2013 to Sep. 2013 & 64 & 36 & 17:00 $-$ 18:25  & --29:30 $-$ --23:30  \\
   92   & Dec. 2013 to Apr. 2014 & 48 & 6 & 07:05 $-$ 08:40  & --31:00 $-$ --26:00  \\
   93 & Apr. 2014 to Sep. 2014 & 80 & 36 & 17:05 $-$ 18:30  & --32:30 $-$ --21:30 \\
   94 & Dec. 2015 to Apr. 2015 & 64 & 34 & 07:15 $-$ 08:30 & --33:00 $-$ --22:00 \\
   \hline
  \end{tabular}
  \label{tablog}
\end{table*}

\setcounter{figure}{0}
\begin{figure*}
\centering
\includegraphics[width=\linewidth]{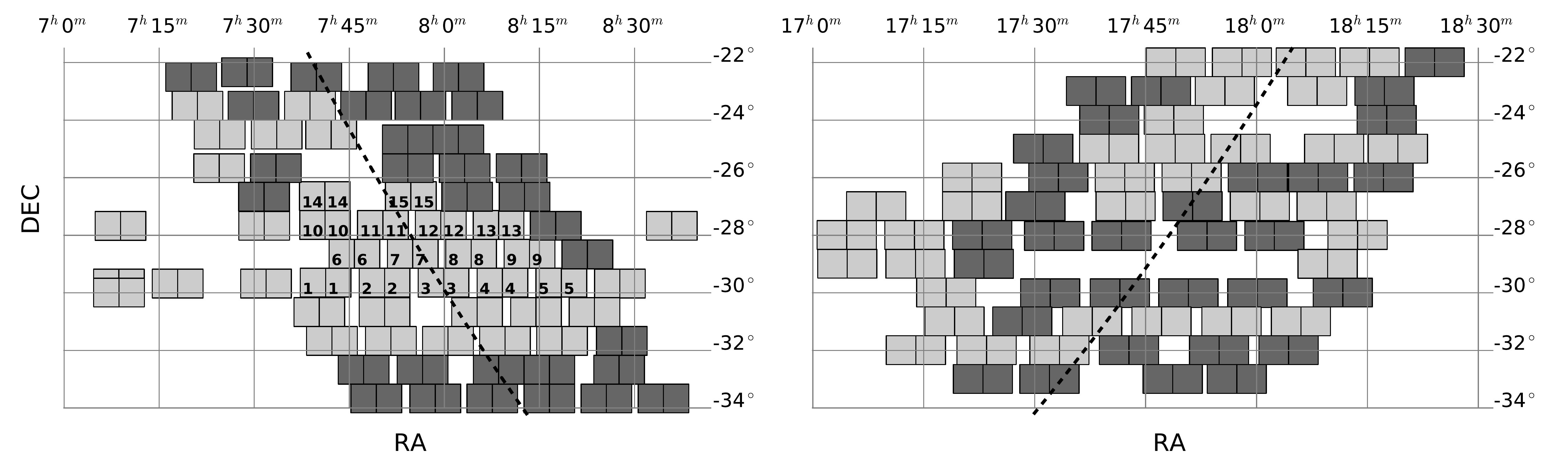}
\caption{The two regions of OmegaWhite field pointings along the Galactic plane (dashed line), including the Galactic Bulge region (right panel).  Fields are either observed (light grey, both panels), scheduled for semester 95 (dark grey, right panel), or scheduled for semester 96 (dark grey, left panel). Semester 88 field pairs are indicated by their identification number, and their respective co-ordinates are listed in Table~\ref{Appfields}.}
\label{Pointings}
\end{figure*}

The OmegaWhite Survey, which is now being conducted using OmegaCam on
the VST, has a similar observing strategy to RATS (See Table~\ref{surveycomp} for a comparison of similar stellar wide-field surveys). However, OmegaWhite aims to cover a much wider area (400 square degrees) at lower average Galactic latitudes ($|b| < 5$ degrees).  In this paper we present: the OmegaWhite observing
strategy, the reduction pipeline, our method for identifying objects of interest, an 
outline of those variables identified in ESO semester 88 (Dec. 2011 --
Apr. 2012), and a discussion of the science goals which can be investigated
using these data.

\section{Observations}
\label{obs}

OmegaWhite observations are taken in $g$-band using the wide-field
instrument OmegaCAM \citep{Kuijken2011} on the VLT Survey Telescope
\citep[VST,][]{Capaccioli2011}. The VST is a 2.65-m wide-field survey
telescope situated at the ESO Paranal Observatory in Northern
Chile. The sole instrument on the telescope, OmegaCAM is a 16k $\times$ 16k
imaging camera, covering one square degree with 32 CCDs (each 2k $\times$ 4k)
at a plate scale of 0.216 arcsec/pixel.

The OmegaWhite survey targets 400 square degrees along the
Galactic plane ($|b|$ \textless\ 5 degrees) and Galactic bulge
($|l|$,$|b|$ \textless\ 10 degrees), of which 148 fields (equivalent
to 148 square degrees) have been observed during ESO semesters
88 to 94 (see Table~\ref{tablog} for semester details, and
Figure~\ref{Pointings} for pointing locations). OmegaWhite fields were
chosen to overlap with the VST Photometric H$\alpha$ Survey of the Southern Galactic Plane \citep[VPHAS+,][]{Drew2014} and the Galactic Bulge Survey \citep[GBS,][]{Jonker2011} pointings in order to obtain broad-band colours and photometric zeropoints, whilst avoiding
bright stars ($V$ \textless\ 5 mag).

\subsection{Observing Strategy}
\label{obsstr}

The OmegaWhite observing strategy has been primarily designed to detect
binary stars with orbital periods in the 5 to 60 minute range,
whilst optimising the compromise between sky coverage and cadence. Two
neighbouring one square degree fields are alternatingly observed in
39 second exposures over an observing duration of 2 hours, with an
observational median cadence of $\sim$2.7 minutes per field. Thus 38
exposures per field are obtained. From the Faint Sky Variability
Survey \citep[FSVS][]{Groot2003}, it was determined that $\sim$25 observations
are necessary in order to reliably identify and accurately determine
the period of short-period variables \citep{Morales2006}. Furthermore,
this exposure time allows for observations to reach magnitude limits
of $g \approx$ 21.5 (10$\sigma$) per exposure. Much deeper than
this is, at times, not useful in the Galactic Plane with seeing-limited
ground-based observations due to stellar crowding. Also, it is unlikely that we will get astrophysically useful identification/follow-up spectra for sources fainter than $g \approx$ 21 mag.

Every two hour observation consists of five observing blocks, each having a respective duration of 21 min, 29 min, 20 min, 21 min, and 29 min (in sequential order). The irregular durations
help break any aliasing that may arise from regular scheduling. At the start of every observing block, the field is reacquired for refocusing and image analysis purposes. For our observations, the seeing is $\leq$ 1\arcsec\ with clear to thin cloud conditions at the start.  However, we are required to relax these constraints for the two observing blocks starting after one hour. Specifically, the seeing is allowed to increase to 2\arcsec\ and thick cloud cover is accepted.

\setcounter{figure}{1}
\begin{figure*}
\centering
\subfloat{\label{FIG:sim}\includegraphics[width=1.0\textwidth]{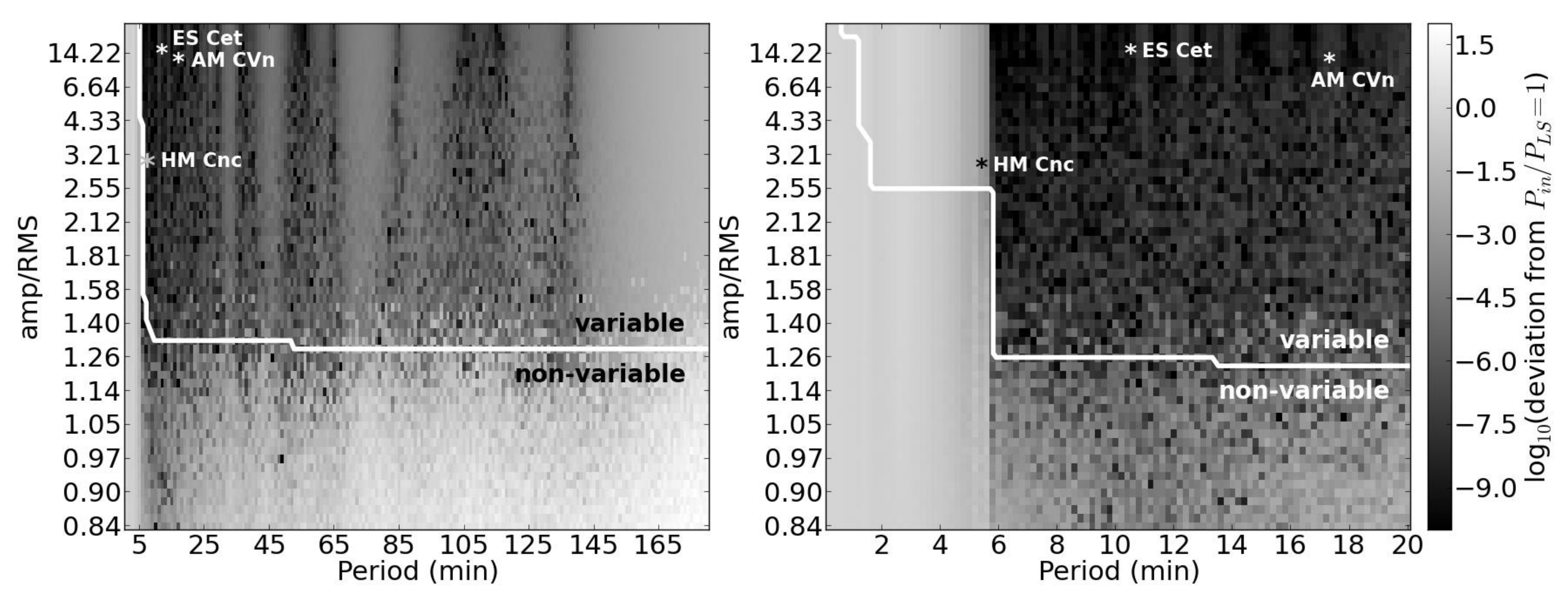}}
\caption{The logarithmic absolute deviation from $P_{in}/P_{LS} = 1$, in period ($P_{in}$) and amp/RMS space, sampled in one minute period bins (left panel), and in 0.2 minute period bins  (right panel). The region above and to the right of the white solid line indicates where sources have  $\log_{10}$(FAP) $< -2.5$, and are therefore considered variable at the 3$\sigma$ level. The location of three short period (HM Cnc, ES Cet, and AM\,CVn) are each indicated by a star in both panels.}
\label{simall}
\end{figure*}

\subsection{Simulations}
\label{simsect}

A key question is whether we can detect the variability
of short-period systems (such as AM\,CVn binaries) using our current
observing strategy and statistical analysis tools. We applied the \textsc{VARTOOLS} suite of software \citep[][used for variability statistics, see Section~\ref{var}]{Hartman2008} to the light curves
of our OmegaWhite sources, allowing us to determine their respective
output Lomb Scargle Period ($P_{LS}$), as well as their false alarm
probability (FAP). A light curve with $\log_{10}$(FAP) $= -2.5$ is
likely to be variable at the 3$\sigma$ confidence level, and the more
negative $\log_{10}$(FAP) is, the more likely it is that the source is
variable (see Section~\ref{var} for more details).

We simulated variable OmegaWhite light curves by applying the actual
observing sequence (as outlined in Section~\ref{obsstr}) to basic
sinusoidal waveforms with increasing input periods ($P_{in}$) in the
range 1 min $\leq P_{in} \leq$ 180 min. To model the intensity of
each exposure, we integrated the sine wave over the corresponding
exposure time interval, to which we then added random Gaussian noise. For the Gaussian sigma level, we used the root mean square (RMS) of the magnitude of a sample of observed light curves, calculating how the standard deviation of the mean for each non-variable light curve in the example field `OW\_88D\_1a' varies as a function of the source's magnitude. We then applied \textsc{VARTOOLS} to the light
curves, allowing us to determine $P_{LS}$ and the FAP value for each
$P_{in}$. Since the probability of finding our target systems depends
on both the variability amplitude of the light curve (amp) and
the RMS (which determines the noise level), we ran our
simulations for a range of RMS to amp ratios (amp/RMS, where 0.01 $<$ amp/RMS $<$ 20.00).

In Figure~\ref{simall}, we show the logarithmic absolute deviation from $P_{in} /
P_{LS} = 1$ for each simulated light curve in amp/RMS and $\log_{10}$(FAP) space. The right panel focuses on the initial 20 minutes of the period range shown in the left panel, and is sampled at a higher rate (in 0.2 minute intervals, as opposed to the 1 minute intervals in the left panel). In both panels, we overlay a cutoff line to indicate the regions where OmegaWhite sources are variable at the 3$\sigma$ level, with $\log_{10}$(FAP) $< -2.5$ (upper right region). 

Therefore, using our observing strategy, we can expect to detect
variable sources if they have properties which have amp/RMS $ \geq$ 1.3 and $P_{in} \geq$ 5.8
min. Variable sources with $P_{in} \geq$ 1.6 min
should be detected provided they have amp/RMS $\geq$ 2.5. Additionally, we found that the absolute deviation from $P_{in} / P_{LS} = 1$ increases for $P_{in} \leq$ 5.8 min and for  $P_{in} \geq$ 140 min. Thus $P_{LS}$ will overestimate $P_{in}$ for sources with $P_{in} \leq$ 5.8 min, and $P_{LS}$ will overestimate $P_{in}$ for sources with $P_{in} \geq$ 140 min.

We applied our observing strategy and \textsc{VARTOOLS} to simulated
light curves of three known short-period AM\,CVn systems (HM Cnc, ES Cet and, AM\,CVn, see Table~\ref{knownamcvn} for system properties). As shown in Figure~\ref{simall}, the resultant $P_{LS}$ and $\log_{10}$(FAP) imply that we should be able to detect all three AM\,CVn systems as significantly variable with $\log_{10}$(FAP) $< -2.5$ (although we will likely overestimate the $P_{orb}$ of HM Cnc).

\setcounter{table}{2}
\begin{table} 
\centering
\caption{Properties of three short-period AM\,CVn systems}
 \begin{tabular}[pos]{c|c|c|c|c|c|}  
  \hline
  AM\,CVn & $P_{orb}$ & $V$ & amp & amp/RMS & Ref. \\
  & (min) & (mag) & (mag) &  &  \\
  \hline
  HM Cnc & 5.36 & 21.1 & 0.150 & 3 & 1,2\\
  ES Cet & 10.3 & 17.1 & 0.075 & 15 & 3 \\
  AM\,CVn & 17.1 & 14.0 & 0.016 & 13 & 4,5,6\\ 
  \hline
  \multicolumn{6}{p{8cm}}{References: (1) \citet{Israel1999}; (2) \citet{Ramsay2002}; (3) \citet{Warner2002}; (4) \citet{Smak1967}; (5) \citet{Warner1972}; (6) \citet{Roelofs2006} }\\
 \end{tabular}
\label{knownamcvn}
\end{table}

\setcounter{figure}{2}
\begin{figure*}
\centering
\subfloat[Reference subframe]{\label{FIG:refim}\includegraphics[width=0.49\textwidth]{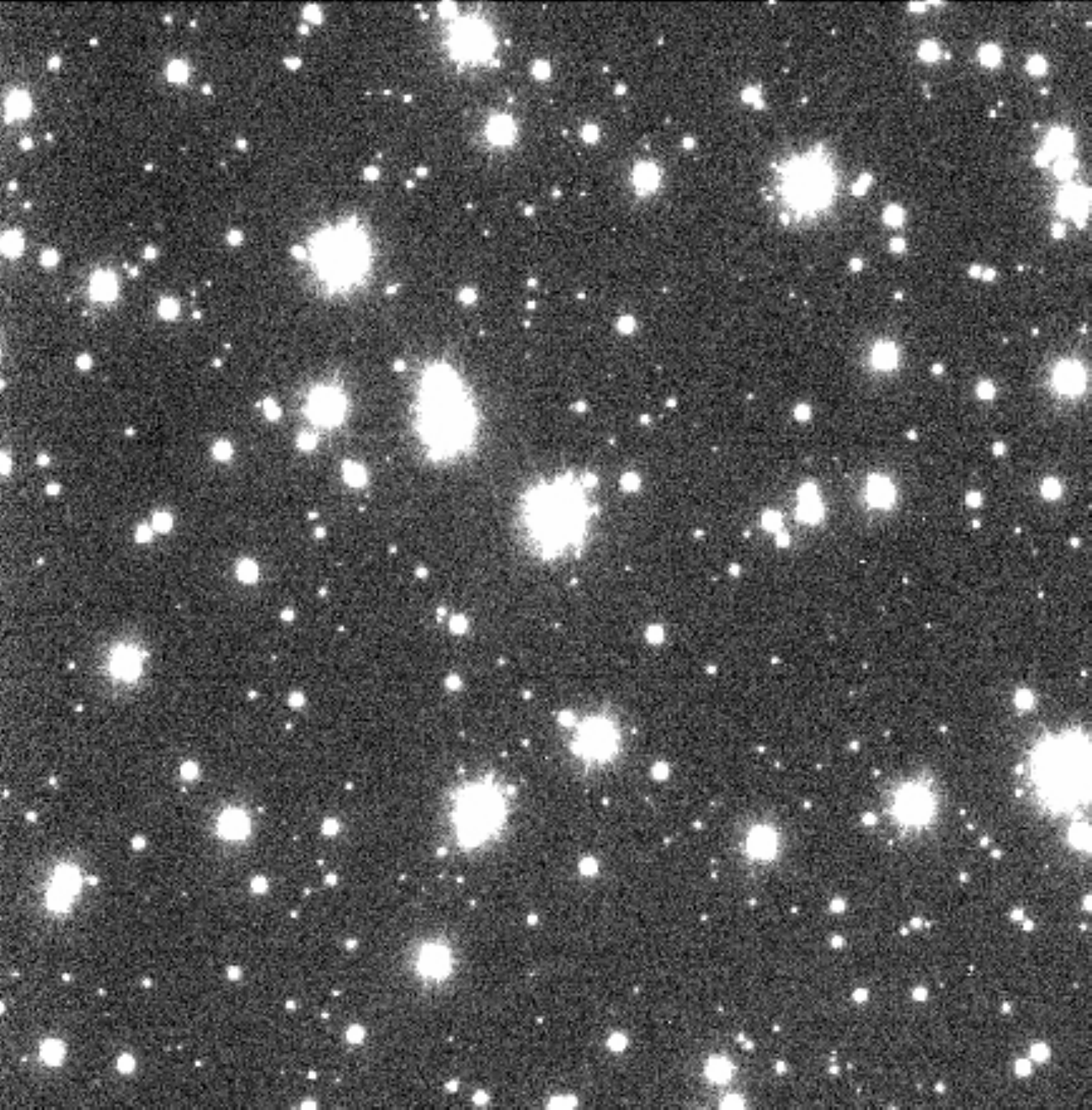}}
\hspace{0.2cm}
\subfloat[Difference subframe]{\label{FIG:difim}\includegraphics[width=0.49\textwidth]{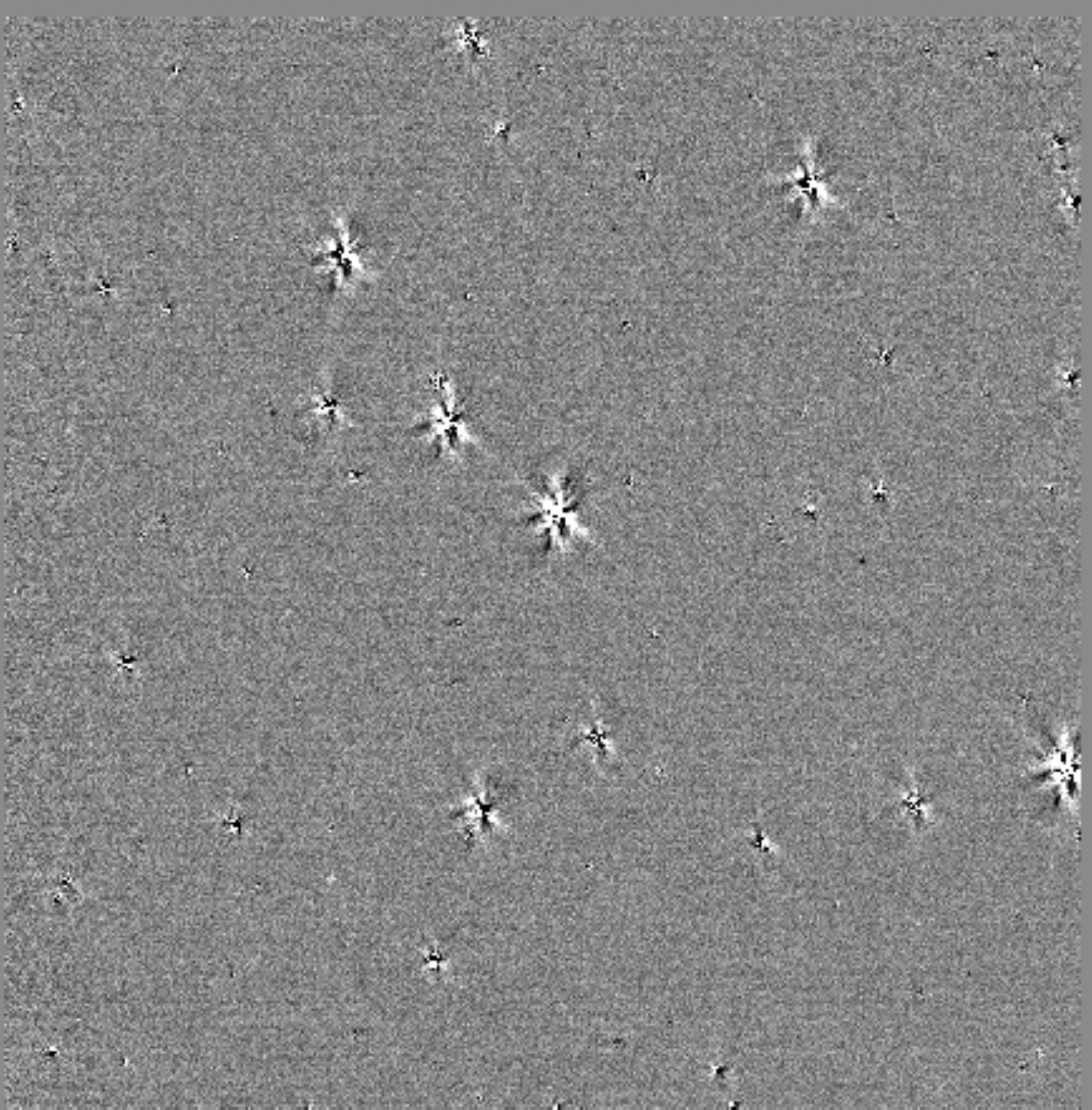}} 
\caption{Difference subframe (right panel) obtained after the
  reference subframe (left panel) is convolved and then subtracted
  from an individual science subframe. Each subframe is 3.75$\arcmin
  \times$ 3.75$\arcmin$. The residuals visible on the difference
    frame were caused by saturated stars.}
\label{seq}
\end{figure*}

\section{Data Analysis}

Our pipeline for cleaning the raw data from instrumental effects,
extracting the light curves and identifying variable stars is a
modified version of that used by the RATS \citep{Barclay2011}
and RATS-Kepler \citep{Ramsay2014} surveys. Here we give an
overview of our procedure.

\subsection{Image reduction and astrometric calibration}

Median bias frames were created for each night, and a median $g$-band
flat field, made using twilight fields, was created for every month
for which science images were available (because flat field data were not
acquired for every night). The raw science images were then cleaned using
standard procedures through bias subtraction and flat field
division. We notice the `negative cross-talk' effect on frames from some chips - mostly on Chip \#96 \citep{Kuijken2011, Drew2014}.

Until early 2014, the VST has suffered various technical problems which caused {the appearance of} scattered light on many images. These issues were only resolved once baffles had been fit to the
telescope (see Drew et al. 2014 for more details). On semester 88 data, we
also found that the autoguiding did not achieve sufficient precision
to ensure that stars were kept fixed on the same set of pixels over
the course of observations. This problem, combined with the potentially
variable sky background, would degrade the resulting photometry. Therefore, we expect that the photometry derived from semester 94 and onwards (see Table~\ref{tablog}) would be significantly improved compared to the results from earlier semesters.

World coordinates were embedded using {\tt Astrometry.net} software
\citep{Lang2010}. This approach matches the sky position of stars in
an image with a set of reference `index' files created using data from the Two Micron All Sky Survey \citep[2MASS,][]{Jarrett2000}, which are split according to the field's
position in the sky and the known angular scale of the image. Initially, we
used the software default index files. However, in crowded stellar fields (as we can find in our data), the astrometric solution led to discrepancies of several arcseconds between the fitted and known positions of 2MASS, or the software failed to find a solution for images of some chips. We sought to solve this problem by setting a brighter magnitude limit on the 2MASS reference stars, as brighter objects are less affected by crowding with faint sources. To implement this solution, we created our customised index files using stars within
the field of OmegaCam, as well as 2MASS data (the default angular
extent of the 2MASS index files is much greater than the field of 
view of OmegaWhite). Additionally, we filtered the 2MASS data to 
include only stars whose $J$-mag photometry has a signal-to-noise ratio greater than five, giving a limit of $J\sim$ 16.5 mag (whereas the default files tend to go deeper). Our approach led to improved astrometric solutions for all chips with a typical residual of $\sigma\sim0.1$\arcsec\ in right ascension (RA) and declination (DEC) when compared to the 2MASS positions. Nevertheless, for a small number of fields which are particularly crowded, the residuals can be an order of magnitude greater. 

\subsection{Light curve creation}
\label{varcreation}

\setcounter{figure}{3}
\begin{figure}
 \centering
 \includegraphics[width=\linewidth]{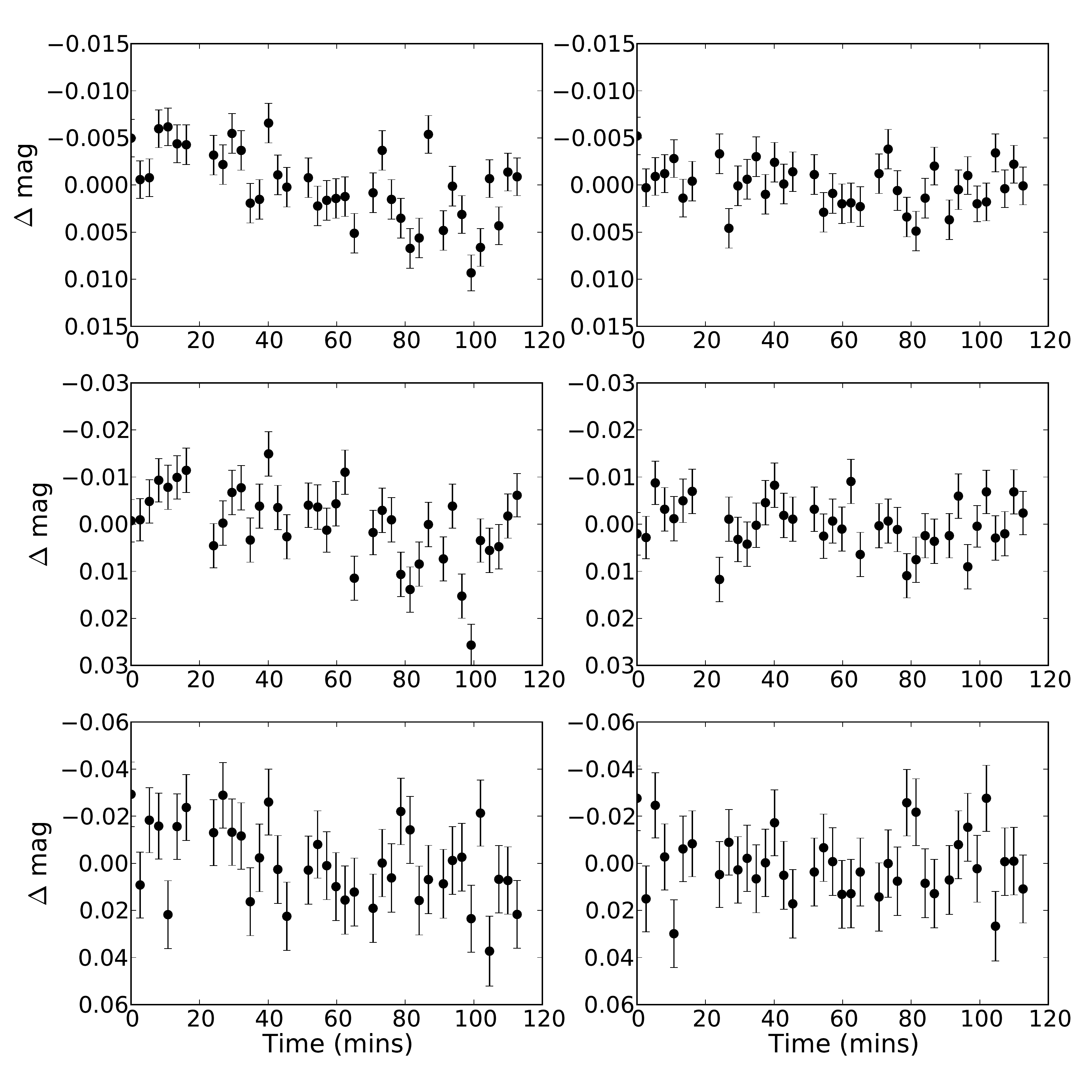}
 \caption{The results of applying the detrending algorithm for
   three light curves (from top to bottom having $g\sim$ 15.3 mag, $g\sim$ 17.0
  mag, and $g\sim$ 18.8 mag). 
   Systematic trends that appear as large-scale wavelength-dependent
   effects tilting the light curves in the left panels are
   successfully corrected for as shown in the right panels.} 
 \label{FIG:ex-syst-trends}
\end{figure}

Since many of our fields are expected to be crowded, we used the {\sl
  difference imaging analysis} technique to measure stellar fluxes, as
implemented in the Difference Image Analysis Package {\tt diapl2}  \citep{Wozniak2000}, which is
an adaptation of the original algorithm outlined in
\citet{Alard1998}. The basic principle consists
of subtracting a reference image from each science frame, after this
reference image is degraded with a convolution kernel (a matrix used
to smooth the seeing of the reference image) to match the seeing of
each individual image. There are known to be mild
variations in the Point Spread Function (PSF) of stars across the detector
\citep{Kuijken2011}. To account for this potential variation, we split
each image into halves before applying the difference
imaging algorithm (splitting the images into smaller units, such as eighths, causes the pipeline to fail in finding suitable stars to create a model PSF in the case of some subframes). An example of a reference subframe and the corresponding difference subframe obtained as the result of subtraction 
are illustrated in Figure~\ref{seq} (Field OW\_88D\_15a, Chip \#65). 

Afterwards, the flux of `residuals' on the subtracted
images is measured through aperture photometry. These residuals are
then added to the flux previously measured on the
reference image. Thus, we obtain one photometric point for each star
per image. Light curves which had less than ten photometric points
were filtered out at this stage. Ultimately, the mean values of the 
measured instrumental magnitudes are calibrated using the AAVSO Photometric All-Sky Survey catalogue \citep[APASS,][]{Henden2009} for reference. Thus, all magnitude values listed in this paper are expressed using the Vega system. \\

The light curves were then corrected for systematic effects in the data
using the {\tt SYSREM} algorithm \citep{Tamuz2005}, which assumes that systematic trends affect measurements in a manner analogous to the effects of atmospheric extinction (dependent on the colour of each star and airmass of each image). The algorithm minimises the global expression: 
\begin{equation}
S^2=\sum_{ij}\frac{(r_{ij}-c_ia_j)^2}{\sigma^2_{ij}},
\end{equation}
where $r_{ij}$ is the residual value of magnitude for each data point, $c_i$ is the colour term for each star $i$, $a_j$ is the airmass for each image $j$ and $\sigma_{ij}$ is the error. Basically, {\tt SYSREM} subtracts the product $c_ia_j$ from the magnitude of each data point  \citep[see][for details]{Tamuz2005,Barclay2011}. In Figure~\ref{FIG:ex-syst-trends}, we show non-detrended (left panels) and detrended (right panels) light curves for three non-variable stars, selected from the same field, OW\_88D\_13a. Their magnitudes are $g\sim$ 15.3 mag, $g\sim$ 17.0 mag, and $g\sim$ 18.8 mag. As shown in Figure~\ref{FIG:ex-syst-trends}, {\tt SYSREM} is able to successfully remove large-scale trends (appearing as tilts in the light curve). These trends are typically the result of
atmospheric extinction effects, and are therefore wavelength-dependent. 

\setcounter{figure}{4}
\begin{figure}
\centering
\subfloat{\label{p881a}\includegraphics[width=0.49\textwidth]{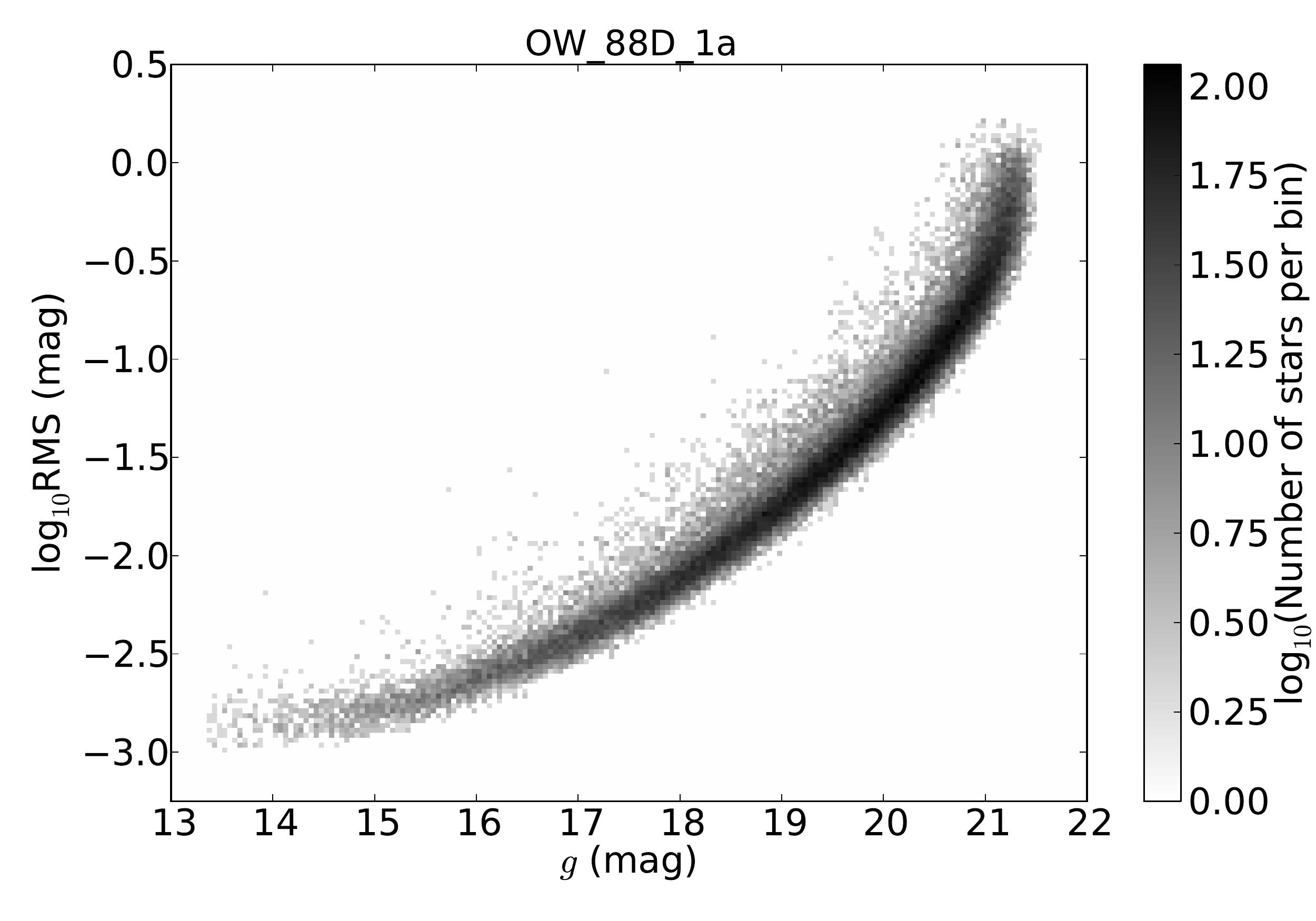}}
\hspace{0.2cm}
\subfloat{\label{p8813a}\includegraphics[width=0.49\textwidth]{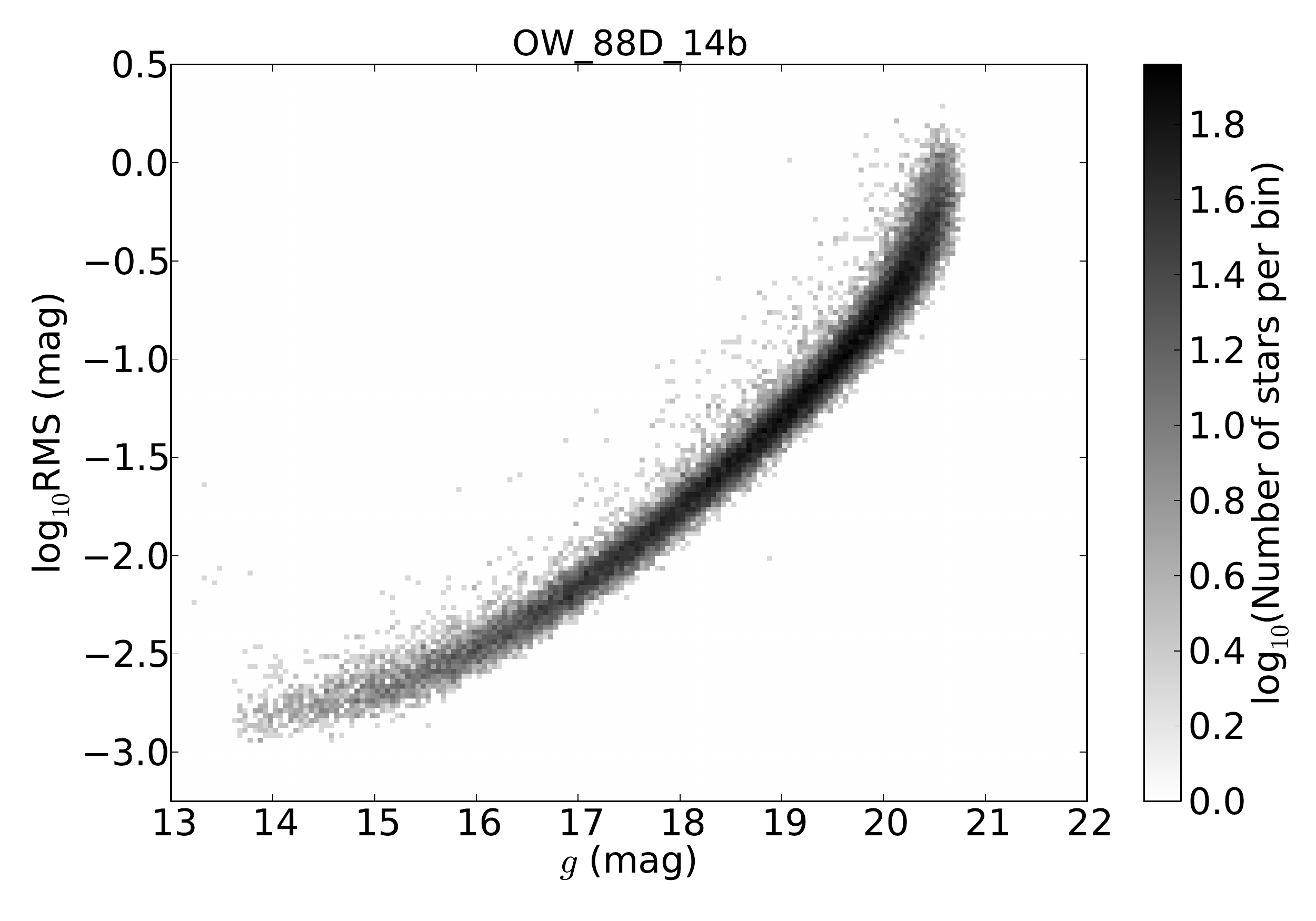}} 
\caption{Distributions of standard deviations of magnitude for each light curve as a function of measured $g$-band magnitude, obtained using data from two fields observed during semester 88. The grey scale represents a 2D histogram showing the number of stars in bins of 0.05 $\times$ 0.025 mags.}
\label{FIG:rmsplots}
\end{figure}

To assess the quality of our photometric measurements, we study the
distribution of the standard deviation from the mean of measured
$g$-band magnitude, for each light curve on a field-by-field and
chip-by-chip base. In Figure~\ref{FIG:rmsplots} we show examples of
such plots for two fields, namely, OW\_88D\_1a and OW\_88D\_14b. 
The RMS plots for almost all of the fields from semester 88 look similar, as expected. However, exceptions are noticed in the case of two particular fields, namely
OW\_88D\_6a and OW\_88D\_6b, which both show a higher RMS scattering over the
entire distribution which we attribute to poor
observing conditions (variable seeing and thick clouds were present
during the observation run).

\subsection{Flagging method for poor photometry detections}
\label{flagging}

In order to investigate the number of bona fide variable stars, some of the 
automatically selected sources were manually checked through
visual inspection of their light curves, their Discrete Fourier Transform (DFT) power spectra and
the corresponding CCD images. As a result of this verification stage, we found that the light curves of many of our detections are affected by poor photometry. In most cases they were wrongly identified as variable stars since they were either within a few arcseconds of very bright objects, or situated in the diffraction spikes of saturated stars. We therefore modified our pipeline to identify three types of possible poor photometry objects. Firstly, we flagged stars located within 30 pixels to the edge of images, or close to bad columns of pixels. Secondly, we flagged very faint stars with low signal-to-noise ratio ($g >$ 21.5 mag) and very bright stars which could be saturated ($g <$ 13.5 mag). Finally, stars with a high background sky were flagged, including the faint sources on diffraction spikes of bright stars. The flagging threshold was set to a value of 1.5$\sigma$ above the sky background median for the entire image.

In Table~\ref{flagged1a}, we note the total number of stars in Field 1a and the number of stars which were flagged for one or more of the three reasons. For example, a very faint star with magnitude $g >$ 21.5 mag and which is close to diffraction spikes would have been flagged twice. In total, 11.9 per cent of the stars in Field 1a were flagged.
  
\setcounter{table}{3}
\begin{table}
\centering
\caption{Results of flagging stars in Field OW\_P88\_1a.  \\
Flag 1 = stars close to edge of images within 30 pixels \\
Flag 2 = saturated and faint stars ($g<$ 13.5 mag and $g>$ 21.5 mag resp.) \\
Flag 3 = stars with high sky background (1.5$\sigma$ above median)}
\begin{tabular}[pos]{|c|c|c|}  
\hline
 & No. Stars & \% \\
\hline
Initial & 60223 & 100 \\ 
Flag 1 & 1926 &  3.2 \\ 
Flag 2 & 3221 &  5.4 \\
Flag 3 & 2274 &  3.8 \\
All flags & 7191 & 11.9 \\
Unflagged & 53032 & 88.1 \\
\hline
\end{tabular}
\label{flagged1a}
\end{table}

\subsection{Variability}
\label{var}

We applied a variety of statistical tests to the light curves using the \textsc{VARTOOLS} Light Curve Analysis Program.  As outlined by \citet{Graham2013} and \citet{Ramsay2014}, different tests are better suited to identifying
particular types of variable stars. For example, the Lomb Scargle test
\citep[LS,][]{Lomb1976,Scargle1982} is suitable for finding
short-period pulsating sources in irregularly spaced data, whilst the
alarm test \citep{Tamuz2006} and Analysis of Variance periodogram
\citep{SchwarzenbergCzerny1989} are ideal for finding eclipsing
binaries and flare stars. Additionally, the Stetson J statistic
\citep{Stetson1996}, which was designed to find Cepheid variables, is
optimally suited to detect high amplitude variables (e.g. contact
binaries, flares and long period pulsators).

Since our main goal is to identify short-period (P$_{orb} < 60$ min)
variable stars, we use the LS periodogram as our main tool. The mean cadence of our observations is 3.2 min, while the median cadence is 2.7 min. Therefore, we chose to set our short period limit to 6.0 min (which corresponds to the Nyquist frequency, also see Figure~\ref{simall}), and set the long period limit to 2.2 hrs (corresponding to the duration of the observations). 
  
For each light curve we identify the period and the
false alarm probability (FAP) corresponding to the most statistically
significant peak in the power spectrum. FAP is a statistic which
describes the probability that the highest peak is caused by random
noise. In Figures~\ref{MADn} and~\ref{FAPall} we show the distribution of all measured 
stars in period - $\log_{10}$(FAP) space, where the period
represents the highest peak in the power spectrum of individual light
curves. The more negative the value of $\log_{10}$(FAP), the higher
the probability that the light curve shows real variability on that
time-scale. In theory, a star having $\log_{10}$(FAP) $= -2.5$ has a
probability of being a variable object at the 3$\sigma$ confidence
level. The vast majority of sources are located in the upper part of
figure, meaning that (as expected) most stars are not detected as variable using our methods.
 Only 0.51 per cent of the 1.6$\times10^{6}$ sources from semester 88 have
$\log_{10}$(FAP) $< -2.5$, i.e. are detected as variable within the 3$\sigma$ confidence interval.

\setcounter{figure}{5}
\begin{figure*}
\centering
\subfloat[Variable stars in Field OW\_88D\_1a.]{\label{MADn}\includegraphics[width=0.79\textwidth]{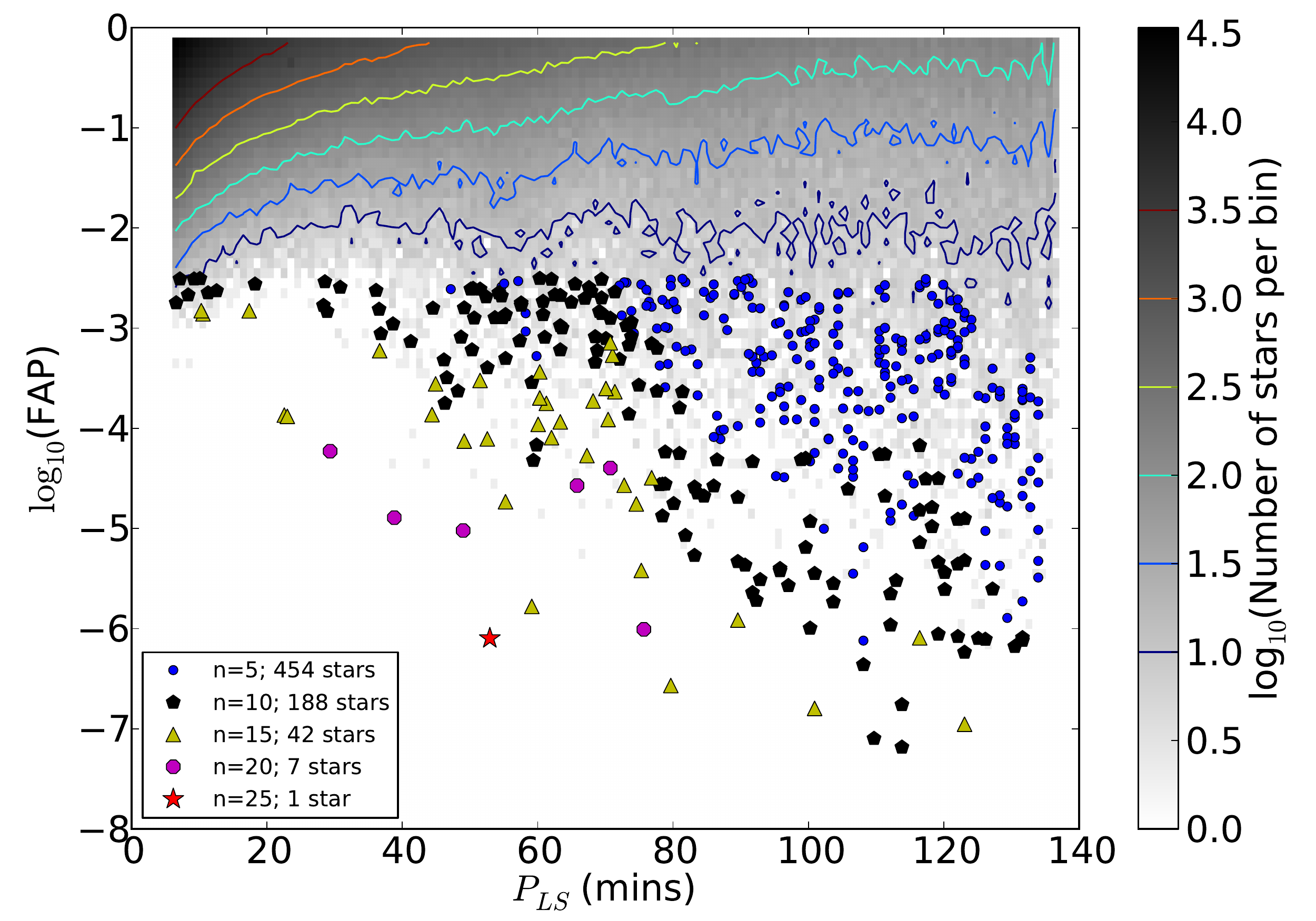}} 
\hspace{0.2cm}
\subfloat[Variable stars in ESO semester 88.]{\label{FAPall}\includegraphics[width=0.79\textwidth]{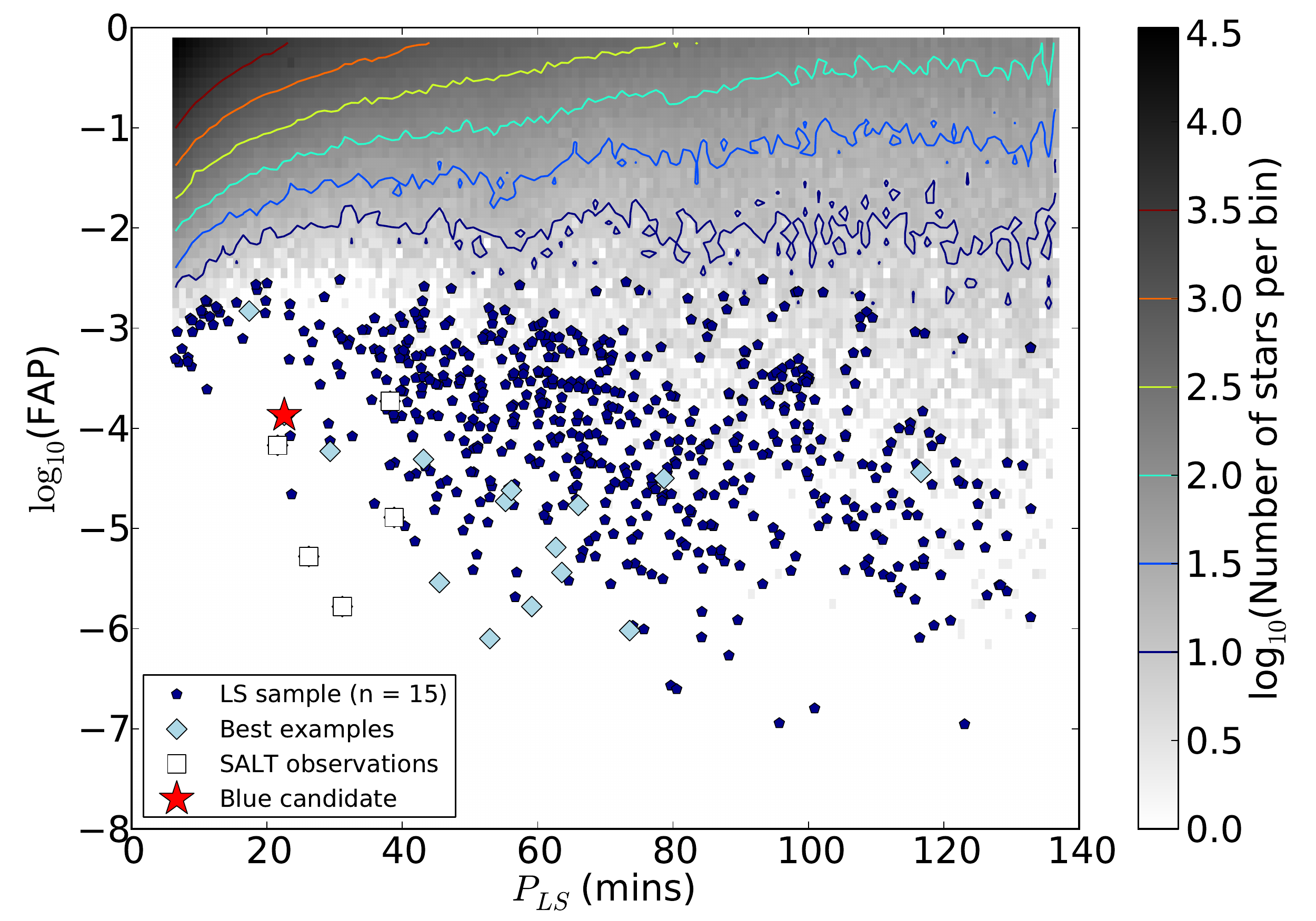}}
\caption{Distribution of all 1.4$\times10^{6}$ unflagged stars from semester 88 data in
  Period - $\log_{10}$(FAP) space. Data are grouped in 1-minute bins on the period axis and in 0.1 units on the $\log_{10}$(FAP) axis. Coloured contours are plotted starting from
  $\log_{10}$(number of stars) = 1 and then at levels incremented by
  0.5 to indicate trends in the distribution. In Figure~\ref{MADn}
  the number of variable candidates automatically selected from Field 1a data
  are plotted on top. In Figure~\ref{FAPall} the variable candidates are overlaid. }
\label{LSMADFAP}
\end{figure*}

However, in the case of real data, the threshold for variability can
be more negative than the theoretical value $\log_{10}$(FAP) $= -2.5$,
because of systematic effects such as red noise. This can be seen in
Figure~\ref{LSMADFAP}, where there is a greater spread in
$\log_{10}$(FAP) values towards longer periods. This indicates the presence of red noise, which affects the stars with possible variability on longer time-scale. Thus, we use the median absolute
deviation from the median (MAD), a more robust statistic to detect the
variable candidates which are outliers (in the very long tail of the
distribution).

We apply the same procedure and codes explained in detail by
\citet{Barclay2011} and \citet{Ramsay2014}. We sort the data according
to period and group them in bins of 2 minutes. For all stars in each
bin, we then compute the median of $\log_{10}$(FAP) values
($median_{LogFAP}$), and the median absolute deviation from
$median_{LogFAP}$ (MAD$_{LogFAP}$), and select as variable candidates
the stars which obey the condition:
\begin{equation}
\centering
\log_{10}(FAP) <  median_{LogFAP} + (n \times MAD_{LogFAP} )
\label{madcondition}
\end{equation}	
where $n$ is an integer which defines how far a source is from the
local median ($median_{LogFAP}$).  Since there is no pre-established
recipe, the stars are selected by experimenting with different values
of $n$.

An example of this experiment is shown in Figure~\ref{MADn}. We overlay the automatically detected variable stars from Field 1a, on top of all stars measured in semester 88. The number of stars selected for each value of $n$ are listed in the legend. As the same stars that were selected for high values of $n$ are also included in the lower $n$ value sets, for clarity in Figure~\ref{MADn} we plot each star only once. We notice that the value of $n$ and the number of selected stars are inversely proportional. This proves that the lower the value of $n$, the higher the probability that the selected stars are poor photometry detections because their light curves show variations caused by systematic effects.
  
We show our sample of variable candidates automatically selected using MAD statistic of LS false alarm probability in Table~\ref{MADvarcand_bonafide}. The total number of automatically selected candidates for a set of values of $n$ are shown in the last column, `Total'.  
Furthermore, we group the stars in
intervals of periods associated with the three AM\,CVn states (based on
disc and outbursting properties). We have chosen the set of variable stars detected for $n$ = 15 in the next steps of data analysis. 

\setcounter{table}{4}
\begin{table} 
\centering
\caption{The variable candidates automatically selected from the entire set of data observed during ESO semester 88, for a set of
  values of $n$ are listed. The selection algorithm was applied on the set of unflagged stars. }
\begin{tabular}[pos]{c|c|c|c|c|}  
\hline
Period & 0-20 min & 20-40 min & P $>$ 40 min & Total   \\
\hline
$n$   & (No. stars)  & (No. stars)  & (No. stars) & (No. stars)  \\
\hline
5  & 177 & 139 & 4546 & 4862 \\
10 & 177 & 136 & 2017 & 2330 \\
15 &  29 &  56 &  528 & 613 \\ 
20 &   0 &  18 &  126 &  144 \\
25 &   0 &   6 &   29 &   35 \\
\hline
\end{tabular}
\label{MADvarcand_bonafide}
\end{table}

\subsection{Colour Analysis}

The VPHAS+ survey \citep{Drew2014} obtains photometric measurements in
$u$-, $g$-, $r$-, $i$- and H$\alpha$-bands along the southern Galactic
plane. Using these multi-band colours, we can identify specific types of
variable stars which lie within certain regions of colour-colour
space. As an example, AM\,CVn systems are
known to be very blue sources and thus occupy a distinctive region of the
$u-g$ vs $g-r$ colour plane \citep{Carter2013}.  We have additionally
cross-matched our data with other multi-colour survey catalogues such
as the Two Micron All Sky Survey \citep[2MASS,][]{Jarrett2000} and the
AAVSO Photometric All-Sky Survey \citep[APASS,][]{Henden2009}. However,
these surveys do not reach the depths of OmegaWhite, and thus some of
our fainter sources currently lack colour information from these surveys.

Although we found that 98 per cent of the stars in Field 1a match stars in four VPHAS+ fields, some of our fields lack colour information because they have not yet been fully covered by the VPHAS+ survey. Out of the 26 deg$^2$ of data observed in semester 88, 7 deg$^2$ (i.e. 26.9 per cent of our fields, namely Fields 4b, 5a, 5b, 8a, 8b, 9a, 9b) lack colour data. Thus, in the case of our sample of 613 variable candidates selected for $n$ = 15 (see Section~\ref{var}), we found information in all three $u$, $g$, $r$ bands for 362 stars (59.1 per cent). We show their distribution in colour - colour space in Figure~\ref{col_col_plots}, and list their properties in Table~\ref{lc_parameters_full} .

\setcounter{figure}{6}
\begin{figure*}
\centering
\subfloat[colour-colour plot of all fields observed during semester 88]{\label{colall}\includegraphics[width=0.75\textwidth]{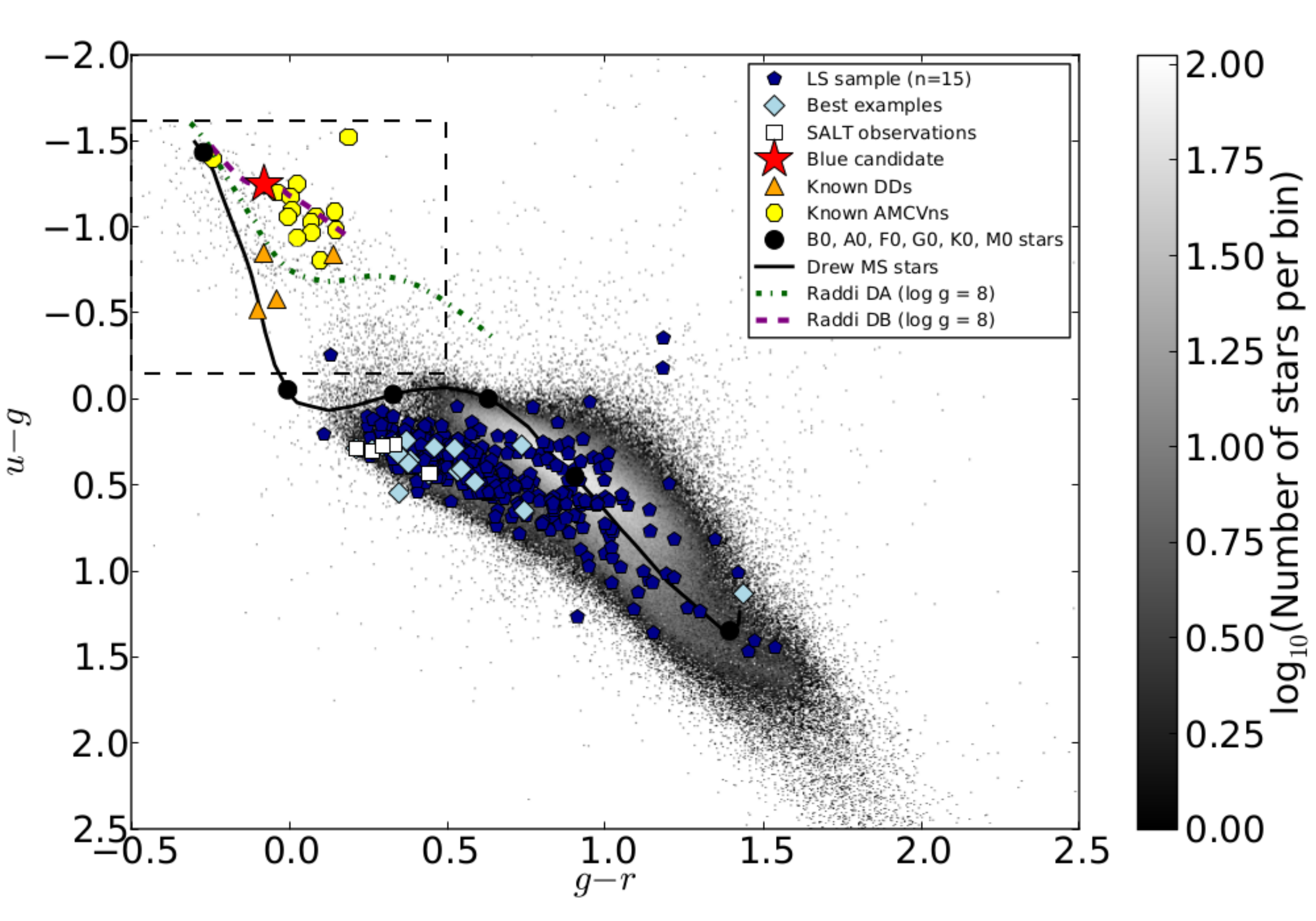}}
\hspace{0.2cm}
\subfloat[Zoom in of the blue region]{\label{colbluesq}\includegraphics[width=0.75\textwidth]{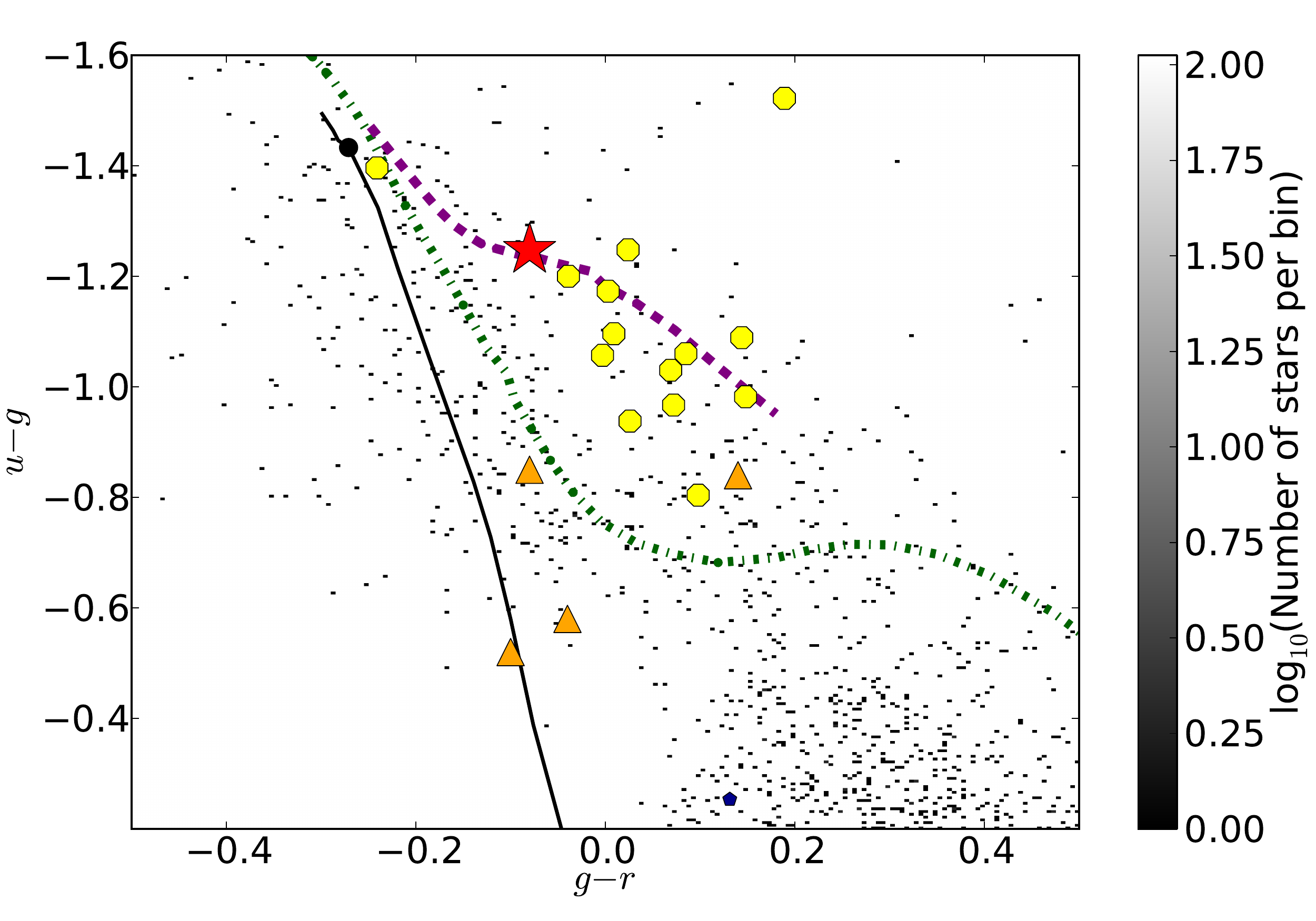}}  
\caption{Distribution of all stars measured in semester 88 data, for which VPHAS+ colour information was available is shown in $u-g$ vs $g-r$ space.  In the upper panel, stars are distributed in a 2D histogram with bins of 0.005 mag units. The sample of variable candidates selected for $n$ = 15 are overlaid. Of the 613 variable candidates, only 362 have both colour indices and thus could be represented. For comparison, we plot 14 of the known AM CVn stars identified in SDSS data as circles \citep{Carter2013}, and the four known detached double white dwarf systems with periods less than 1 hour as triangles \citep{Kilic2011,Hermes2012,Kilic2014}. We also plot tracks for main sequence stars, DA and DB white dwarfs. For the MS track, we used the synthetic colours provided by \citet{Drew2014} for R$_v$=3.1 and extinction coefficient A$_0$ = 0. The synthetic colours for the cooling tracks were taken from Raddi et al. (in prep, private communication) for models with a surface gravity of log $g$ = 8.0. A close-up of the blue-square (i.e. the square where all the blue UCBs are expected to be found) is shown in the bottom panel. One of our variable sources, OW J074106.1--294811.0, lies in this blue region together with other known AM CVn stars.}
\label{col_col_plots}
\end{figure*}

\section{Initial Survey Results}
\label{results}

\setcounter{figure}{7}
\begin{figure*}
\centering
\includegraphics[width=0.95\textwidth]{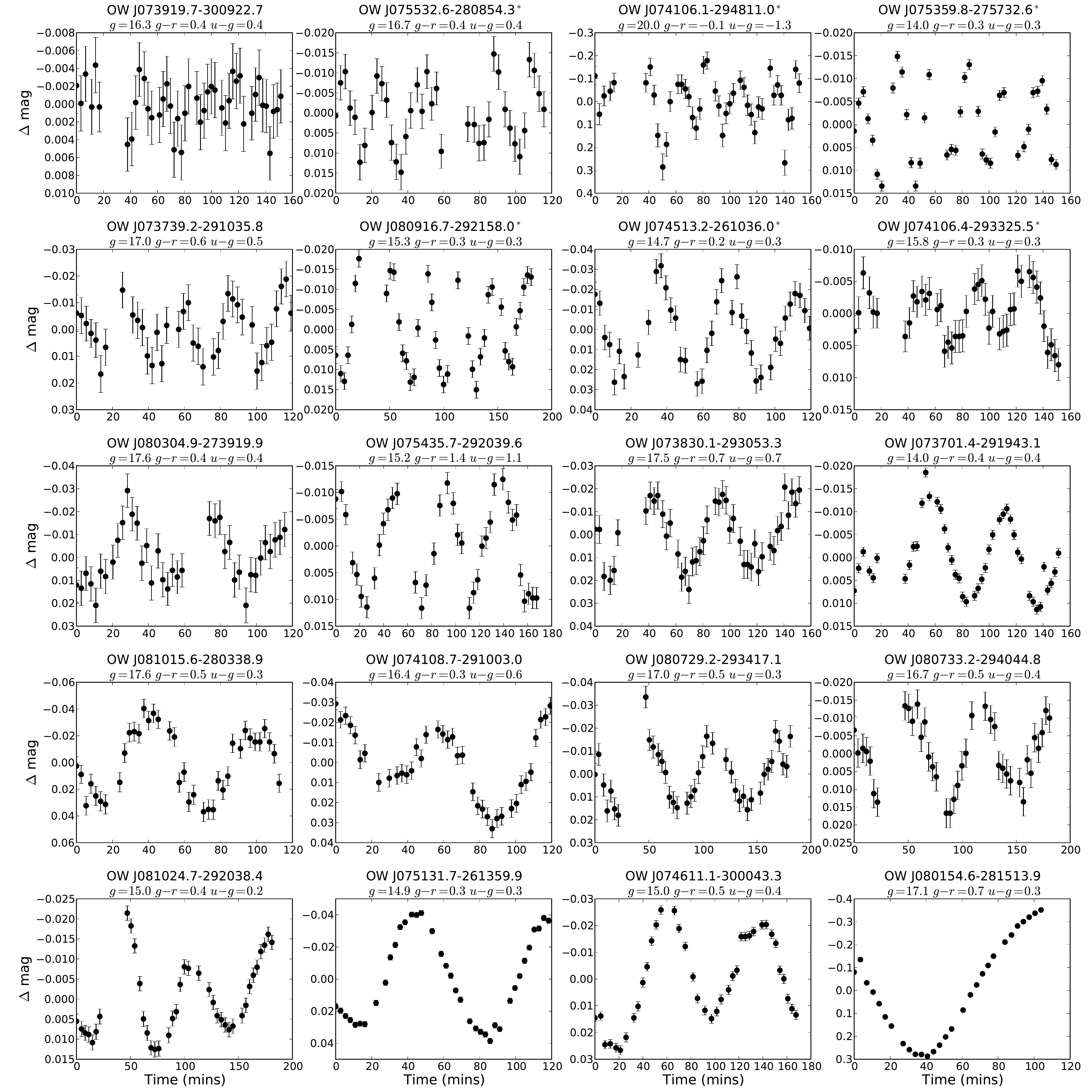}
\caption{Example set of light curves selected from 613 variable candidates detected in semester 88 data. The six stars marked with an asterisk have been followed-up spectroscopically using the RSS spectrograph on SALT. }
\label{P88LCs}
\end{figure*}

We have extracted the light curves of 1.6$\times10^{6}$ stars from the initial 26 square degrees of the OmegaWhite Survey (semester 88 data). Of these stars, 2.0$\times10^{5}$ (12.5 per cent) were flagged in the manner described in Section~\ref{flagging}. Variable candidates were selected from the 1.4 $\times10^{6}$ unflagged stars remaining (87.5 per cent), using the automatic method described in Section~\ref{var}. We thus identified a subset of 613 sources that were significantly variable (with  $\log_{10}$(FAP) $< -2.5$) using $n$ = 15 in our MAD routines, and list their properties in Table~\ref{lc_parameters_full}. Of these, 29 sources exhibit dominant periods less than 20 minutes (4.7 per cent), 56 sources have periods between 20 minutes and 40 minutes (9.1 per cent), and 528 sources have periods greater than 40 minutes (86.2 per cent). 

Furthermore, a subsample of 20 variable stars with both colour indices available has been selected as the most significant examples, after inspecting the light curves, DFT power spectra, and CCD reference images of each source. They were selected from the sample for $n$ = 20, and from the 29 stars selected for $n$ = 15 for periods smaller than 20 minutes. The individual light curves of these examples are displayed in Figure~\ref{P88LCs} and their properties are listed in Table~\ref{lc_parameters}. Only the first star shows sinusoidal modulations on a period smaller than 20 minutes, seven have periods between 20 - 40 minutes, and the rest exhibit variations on longer periods.

In Figure~\ref{col_col_plots}, we show the colour indices of 1.1 $\times10^{6}$ unflagged  
colour-matched stars in the $u-g$ vs $g-r$ plane. We overlay the colours of the 362 variable candidates which show colour information and for comparison plot 14 of the known AM\,CVn stars identified in SDSS data \citep{Carter2013}, as well as the four known detached double white dwarf systems with
periods less than 1 hour \citep{Kilic2011,Hermes2012,Kilic2014}.  One of the 362 variables, namely OW\,J074106.1--294811.2 ($g-r = -0.08$ and $u-g = -1.25$) has colours consistent with those of an AM\,CVn star, a short-period double-detached system, or a pulsating white-dwarf/subdwarf candidate (as shown in Figure~\ref{col_col_plots} and in Table~\ref{lc_parameters}). We present follow-up spectral observations of this source, and discuss its possible nature in Section~\ref{followup}.

The majority of the 613 variable candidates exhibit periodic variations, amplitudes or colours that are more consistent with pulsating sources, and are likely to be either $\delta$ Sct stars, SX Phe pulsators or pulsating white dwarfs. We have also detected longer period systems, such as contact binary candidates. An example of such a system with P$_{LS} \ga$ 116 min is shown in Figure~\ref{P88LCs}.

\setcounter{table}{5}
\begin{table*}
\centering
\caption{{Parameters of the variable stars shown in
  Figure~\ref{P88LCs}: star ID; RA and Dec; Lomb Scargle period 
  (P$_{LS}$) and false alarm probability ($\log_{10}$(FAP)) of the highest peak in the periodogram, OW calibrated $g$-band magnitude, RMS of magnitude, VPHAS+ colour indices $g-r$ and $u-g$, and
  comments on variability type. }}
\begin{tabular}[pos]{|l|c|c|r|c|c|c|r|r|c|c|} 
\hline
 Star ID & RA (J2000) & DEC (J2000) & $P_{LS}$ & $\log_{10}$(FAP) & OW$g$ & RMS & $g-r$ & $u-g$ & Comments \\
 & (hh:mm:ss) & (\degr:\arcmin:\arcsec) &  (min) &  & (mag) & (mag) & (mag) & (mag) &  &  \\
\hline
OW J073919.7--300922.7 & 07:39:19.7 & --30:09:22.7 & 17.38 & --2.83 & 16.26 & 0.002 & 0.38 & 0.37 & \\
OW J075532.6--280854.3$^{*}$ & 07:55:32.6 & --28:08:54.3 & 21.57 & --4.17 & 16.70 & 0.008 & 0.44 & 0.44 & $\delta$ Sct (F5V) \\
OW J074106.1--294811.0$^{*}$ & 07:41:06.1 & --29:48:11.0 & 22.56 & --3.87 & 20.02 & 0.106 & --0.08 & --1.25 & pulsating subdwarf?  \\
OW J075359.8--275732.6$^{*}$ & 07:53:59.8 & --27:57:32.6 & 26.19 & --5.28 & 13.98 & 0.008 & 0.26 & 0.30 & $\delta$ Sct (F2V)  \\
OW J073739.2--291035.8 & 07:37:39.2 & --29:10:35.8 & 29.34 & --4.23 & 16.97 & 0.006 & 0.59 & 0.48 & \\
OW J080916.7--292158.0$^{*}$ & 08:09:16.7 & --29:21:58.0 & 31.14 & --5.78 & 15.30 & 0.010 & 0.29 & 0.27 &  $\delta$ Sct (F2V) \\
OW J074513.2--261036.0$^{*}$ & 07:45:13.2 & --26:10:36.0 & 38.19 & --3.73 & 14.73 & 0.004 & 0.21 & 0.29 & $\delta$ Sct (F0V)  \\
OW J074106.4--293325.5$^{*}$ & 07:41:06.4 & --29:33:25.5 & 38.81 & --4.89 & 15.75 & 0.004 & 0.33 & 0.27 & $\delta$ Sct (F2V)  \\
OW J080304.9--273919.9 & 08:03:04.9 & --27:39:19.9 & 43.13 & --4.31 & 17.61 & 0.012 & 0.38 & 0.36 & \\
OW J075435.7--292039.6 & 07:54:35.7 & --29:20:39.6 & 45.49 & --5.54 & 15.23 & 0.008 & 1.44 & 1.13 & \\
OW J073830.1--293053.3 & 07:38:30.1 & --29:30:53.3 & 52.93 & --6.10 & 17.47 & 0.012 & 0.74 & 0.65 & \\
OW J073701.4--291943.1 & 07:37:01.4 & --29:19:43.1 & 55.25 & --4.73 & 14.02 & 0.007 & 0.38 & 0.37 & \\
OW J081015.6--280338.9 & 08:10:15.6 & --28:03:38.9 & 56.12 & --4.62 & 17.62 & 0.021 & 0.52 & 0.29 & \\
OW J074108.7--291003.0 & 07:41:08.7 & --29:10:03.0 & 59.13 & --5.78 & 16.41 & 0.012 & 0.35 & 0.55 & \\
OW J080729.2--293417.1 & 08:07:29.2 & --29:34:17.1 & 62.68 & --5.19 & 17.00 & 0.012 & 0.46 & 0.28 & \\
OW J080733.2--294044.8 & 08:07:33.2 & --29:40:44.8 & 63.57 & --5.44 & 16.71 & 0.009 & 0.53 & 0.42 & \\
OW J081024.7--292038.4 & 08:10:24.7 & --29:20:38.4 & 66.01 & --4.77 & 15.04 & 0.009 & 0.37 & 0.24 & \\
OW J075131.7--261359.9 & 07:51:31.7 & --26:13:59.9 & 73.59 & --6.02 & 14.89 & 0.027 & 0.34 & 0.32 & \\
OW J074611.1--300043.3 & 07:46:11.1 & --30:00:43.3 & 78.67 & --4.50 & 15.01 & 0.012 & 0.54 & 0.41 & \\
OW J080154.6--281513.9 & 08:01:54.6 & --28:15:13.9 & 116.63 & --4.44 & 17.11 & 0.210 & 0.74 & 0.27 & Contact binary \\
\hline
\multicolumn{10}{p{12cm}}{$^{*}$Observed spectroscopically with SALT, results presented in Section~\ref{followup} }\\
\end{tabular}
\label{lc_parameters}
\end{table*}

\subsection{Follow-up Spectroscopic Observations}
\label{followup}

\setcounter{table}{6}
\begin{table*} 
\centering
\caption{Approximate spectral type of SALT-observed variables}
\begin{tabular}{|c|c|c|c|c|} 
\hline
SALT-observed& EW (Na \textsc{I}) & EW (H$\alpha$) & EW Ratio & $\sim$ Spectral  \\
Variable& &  &  (Na \textsc{I}/H$\alpha$) & Type  \\
\hline 
OW\,J075532.6--280854.3 & 3.033 $\pm$  0.082 & 15.599 $\pm$  0.133  & 0.194 $\pm$ 0.028 & F5V \\
OW\,J074106.4--293325.5 & 1.948 $\pm$  0.030 & 14.443 $\pm$  0.193  & 0.135 $\pm$ 0.020 & F2V \\
OW\,J080916.7--292158.0 & 1.938 $\pm$  0.019 & 15.971 $\pm$  0.201  & 0.121 $\pm$ 0.016 & F2V \\
OW\,J075359.8--275732.6 & 1.661 $\pm$  0.022 & 15.775 $\pm$  0.199  & 0.105 $\pm$ 0.018 & F2V \\
OW\,J074513.2--261036.0  & 1.071 $\pm$  0.008 & 16.297 $\pm$  0.068 & 0.065 $\pm$ 0.066 & F0V \\
\hline
\end{tabular}
\label{spectraltype}
\end{table*}

We have so far obtained identification spectroscopy for six variable
candidates (OW\,J080916.7--292158.0, OW\,J075359.8--275732.6,
OW\,J075532.6--280854.3, OW\,J074513.2--261036.0, OW\,J074106.4--293325.5, and OW\,J074106.1--294811.0 )
using the 10-m Southern African Large Telescope \citep{Buckley2006}
between December 2014 and April 2015. These variables were selected
for follow-up observations as they were some of our most likely UCB candidates: they exhibited relatively fast variations with low $\log_{10}$(FAP) values (see Figure~\ref{MADn}), and are some of the
bluest sources of our 362 variable candidates. Of these targets, OW\,J074106.1--294811.2 is the only target to exhibit colours consistent with AM\,CVn systems, as shown in Figure~\ref{col_col_plots}). 

Three exposures of 180 s each were obtained for five of the brighter targets (as listed in Table~\ref{spectraltype}) using the Robert Stobie Spectrograph \citep[RSS,][]{Kobulnicky2003} in long-slit mode, with a 1.5\arcsec\ slit and the PG0300 grating (providing a dispersion of
3.04 \AA\ per binned pixel and a wavelength range of $\sim$3200 - 9000 \AA). For our faint blue target OW\,J074106.1--294811.0, we obtained a set of two exposures of 1200 s each over two nights (for a total of four 1200s exposures). Spectra were reduced using standard \textsc{IRAF} \citep{Tody1986,Tody1993} and \textsc{PYSALT} \citep{Crawford2014} routines, with either an Ar arc or a Xe arc for wavelength calibration. The resulting spectral resolution was $\sim$4 \AA\ for the mean combined spectrum of each
source. 

The spectra for each of the five brighter variables appear to be similar; they all exhibit
strong Balmer absorption lines with some evidence of weaker Na \textsc{I} and
Mg \textsc{I} lines. In order to identify the spectral type, we measured the
equivalent width (EW) ratio of the Na \textsc{I} doublet (5896 \AA, 5890 \AA) to H$\alpha$ for each variable, and compared these values with the EW ratios measured in various spectral types from model {\tt Atlas9} data \citep{Castelli2004}. In Figure~\ref{SALTratio}, we overlay our
variable EW ratios on the measured EW ratios for several spectral
types (fitted by an exponential curve). Hence, we can deduce that our
variables  observed using SALT are most likely to be late A-type or early F-type stars, and thus possible
low-amplitude $\delta$ Sct.

\setcounter{figure}{8}
\begin{figure}
\centering
\includegraphics[width=\linewidth]{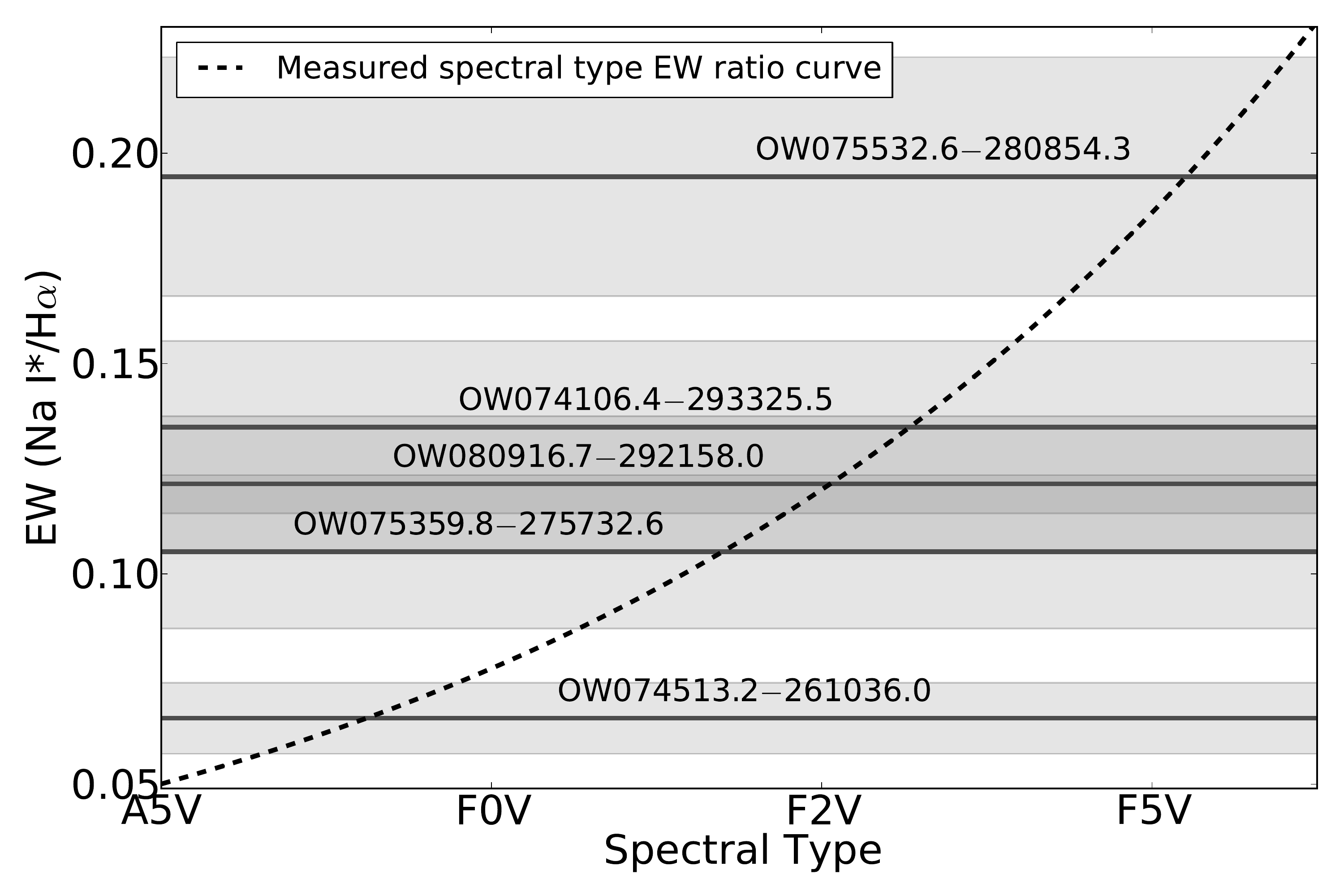}
\caption{Equivalent Width ratios of Na \textsc{I} doublet (5896 \AA,
  5890 \AA) to H$\alpha$ for a range of spectral types. Measured EW ratios increase
  exponentially towards later spectral types (as shown by the dashed
  curved line). The measured EW ratio for each of the five SALT
  variables are shown by a line spanning all spectral types, with
  respective measurement uncertainties indicated by shading. Likely
  spectral classes for each SALT variable are listed in
  Table~\ref{spectraltype}.}
\label{SALTratio}
\end{figure}

\setcounter{figure}{9}
\begin{figure*}
\centering
\includegraphics[width=\linewidth]{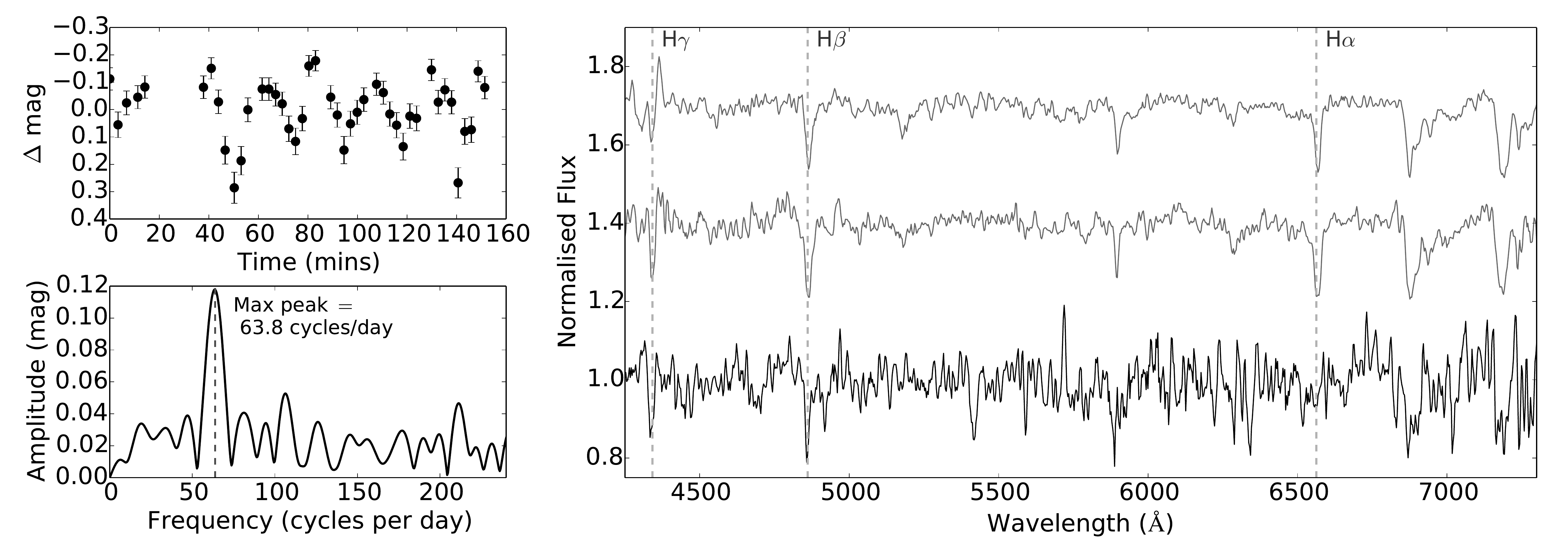}
\caption{Properties of blue variable OW\,J074106.1--294811.0, including the OmegaWhite light curve (upper left panel), the Discrete Fourier Transform power spectrum (lower left panel), and the SALT-observed spectrum (right panel). We plot the continuum-normalised spectra of the target (lower black spectrum) and the spectra of two neighbouring stars for comparison (two upper grey spectra). All spectra have been smoothed by a running average of 3 and an offset has been applied for visualisation purposes.}
\label{bluevarprop}
\end{figure*}

Prior to acquiring SALT spectra, the target OW\,J074106.1--294811.0 was our most likely UCB candidate. This source is relatively faint ($g$ = 20.02 mag ) and the light curve shows sinusoidal modulations on
a period of 22.6 minutes (see Figure~\ref{bluevarprop} for the light curve and Discrete Fourier transform of this target). In the right panel of Figure~\ref{bluevarprop}, we show the SALT-observed spectrum of this target, along with two spectra from neighbouring stars for comparison. In the target spectra, there is weak evidence for the Balmer absorption lines H$\beta$ and H$\gamma$, but no H$\alpha$ (either due to intrinsic weakness of the line, or (partial) infill by emission). Thus, this object is not likely to be a UCB (which would exhibit relatively strong helium emission and little to no hydrogen). Furthermore, the source is relatively blue which is consistent with the properties of a massive hot DA white dwarf. However, the absorption lines are relatively narrow, and thus it is unlikely that this is the case.  The OmegaWhite $g$-band magnitude we observed is $g \approx$ 20.02 mag, while it was recorded by VPHAS+, on two separate epochs, to have $g \approx$ 20.00 mag and $g \approx$ 19.98 mag respectively. This implies that the target is not likely to be a cataclysmic variable  in outburst, where the variations may be due to strong quasi-periodic oscillations (QPOs, see \citet{Morales2002} for examples of outbursting CVs). Although very rare, this does not rule out that the variations may be due to QPOs during quiescence (for e.g. RAT J1953+1859 exhibits high amplitude QPOs during quiescence \citep{Ramsay2009}).  Since it is a very blue source, it is possibly a pulsating subdwarf, although with a relatively long pulsation period. Dedicated high speed photometry of this object should reveal if this object is indeed a pulsator.

\section{Discussion and Summary}

The initial 26 square degrees of OmegaWhite data have been fully processed and the light curves of 1.6$\times10^{6}$ stars have been extracted. Of these stars, 12.5 per cent were flagged as having light curves which could have been degraded by various effects, and were thus disregarded when identifying variable candidates using the verification methods described in Section~\ref{var}. We find that 0.4 per cent of the stars in semester 88 satisfy our selection criteria: they are not flagged as poor photometry detections, their Lomb Scargle periods are between 5 min and 120 min, their $\log_{10}$(FAP) $< -2.5$, and they were selected using a MAD $n$ value $\geq$ 15.  Furthermore, we find that more than 80 per cent (the majority) of the 613 sources identified as variables through our pipeline (using $n$ = 15)  have P$_{LS}$ $>$ 40 min, whereas less than 8 per cent of our variables have P$_{LS}$ $<$ 20 min.  

In Table~\ref{lc_parameters} and Figure~\ref{P88LCs}, we show the properties of 20 relatively blue sources as examples of light curves from our 613 variable candidates. The light curves of 19 of these variables each exhibit variations on a dominant period between 22 min and 80 min. We are also able to detect relatively longer period systems, as illustrated by the contact binary candidate (shown in Figure~\ref{P88LCs}). 

Using spectral identification, we determined that at least five of the short-period variable candidates are likely to be low-amplitude $\delta$ Sct type stars. The remaining 608 variable candidates, including fourteen of the example variables from Table~\ref{lc_parameters}, have not yet been identified with spectroscopy. Some of these variables occupy a similar region of the colour-colour space to our five $\delta$ Sct candidates (as shown in Figure~\ref{col_col_plots}), and also exhibit fast modulations with periods between 17 min and 80 min. In addition to more detailed analysis of the photometry of these systems, spectroscopy is now needed to determine whether the majority are either low-amplitude $\delta$ Sct stars, SX Phe pulsators or pulsating white dwarfs. 

Within the initial 26 square degrees of OmegaWhite, only one of our 613 candidate variables
exhibit the very blue colours consistent with AM\,CVn systems \citep{Carter2013}. However, as it exhibits Balmer absorption lines in its spectrum, it is unlikely to be an UCB. As it is a very blue source, with weak thin hydrogen lines, it is more likely to be a pulsating subdwarf star. Although we have not discovered any AM\,CVn within the initial 26 square degrees, this is a small fraction of our expected coverage and is thus consistent with recent AM\,CVn space density estimates \citep{Carter2013}. Using these estimates, we expect less than one AM\,CVn within the initial 26 square degrees. In a future OmegaWhite paper that will examine $\sim$150 square degrees of the Galactic Plane, we plan to make detailed simulations of the expected AM CVn space densities as a function of Galactic latitude. 

Now that we have formalised our verification procedure and analysed the initial results, we will apply these techniques to future OmegaWhite data. So far 148 square degrees of OmegaWhite data has been
observed, and we have been allocated an additional 64 square degrees (see Section~\ref{obs} for details). This will allow us to get a better understanding of the space density of sources such as UCBs and
$\delta$ Sct along the Galactic plane, as well as within the Galactic Bulge. Furthermore, we are planning photometric and spectroscopic follow-up campaigns for variable candidates, which is essential for their precise classification.

\section{ACKNOWLEDGEMENTS}

The authors gratefully acknowledge funding from the Erasmus Mundus
Programme SAPIENT, the National Research Foundation of South Africa
(NRF), the Nederlandse Organisatie voor Wetenschappelijk Onderzoek
(the Dutch Organisation for Science Research), Radboud University and
the University of Cape Town.  The ESO observations used in this paper are based on observations made with ESO Telescopes at the La Silla Paranal Observatory under programme IDs: 088.D-4010(B), 090.D-0703(A), 090.D-0703(B), 091.D-0716(A), 091.D-0716(B), 092.D-0853(B), 093.D-0937(A), 093.D-0753(A),  094.D-0502(A), 094.D-0502(B), and  177.D-3023 (VPHAS+). Some of the observations reported in
this paper were obtained with the Southern African Large Telescope
(SALT) under program 2014-2-SCI-030 (PI: Sally Macfarlane). Furthermore, we thank the anonymous referee for the useful comments which have helped to improve the paper.

\bibliographystyle{mn2e.bst}
\bibliography{references}

\begin{thebibliography}{}

\bibitem[\protect\citeauthoryear{{Alard} \& {Lupton}}{{Alard} \&
  {Lupton}}{1998}]{Alard1998}
{Alard} C.,  {Lupton} R.~H.,  1998, \apj, 503, 325

\bibitem[\protect\citeauthoryear{{Amaro-Seoane} et~al.,}{{Amaro-Seoane}
  et~al.}{2013}]{Amaro-Seoane2013}
{Amaro-Seoane} P.,  et~al., 2013, GW Notes, Vol.~6, p.~4-110, 6, 4

\bibitem[\protect\citeauthoryear{{Barclay}, {Ramsay}, {Hakala}, {Napiwotzki},
  {Nelemans}, {Potter} \& {Todd}}{{Barclay} et~al.}{2011}]{Barclay2011}
{Barclay} T.,  {Ramsay} G.,  {Hakala} P.,  {Napiwotzki} R.,  {Nelemans} G.,
  {Potter} S.,    {Todd} I.,  2011, \mnras, 413, 2696

\bibitem[\protect\citeauthoryear{{Becker} et~al.,}{{Becker}
  et~al.}{2004}]{Becker2004}
{Becker} A.~C.,  et~al., 2004, \apj, 611, 418

\bibitem[\protect\citeauthoryear{{Borucki} et~al.,}{{Borucki}
  et~al.}{2010}]{Borucki2010}
{Borucki} W.~J.,  et~al., 2010, Science, 327, 977

\bibitem[\protect\citeauthoryear{{Breger}}{{Breger}}{1979}]{Breger1979}
{Breger} M.,  1979, \pasp, 91, 5

\bibitem[\protect\citeauthoryear{{Breger}}{{Breger}}{2000}]{Breger2000}
{Breger} M.,  2000, in {Breger} M.,  {Montgomery} M.,  eds, Delta Scuti and
  Related Stars Vol.~210 of Astronomical Society of the Pacific Conference
  Series, {{$\delta$} Scuti stars (Review)}.
p.~3

\bibitem[\protect\citeauthoryear{{Buckley}, {Swart} \& {Meiring}}{{Buckley}
  et~al.}{2006}]{Buckley2006}
{Buckley} D.~A.~H.,  {Swart} G.~P.,    {Meiring} J.~G.,  2006, in Society of
  Photo-Optical Instrumentation Engineers (SPIE) Conference Series Vol.~6267 of
  Society of Photo-Optical Instrumentation Engineers (SPIE) Conference Series,
  {Completion and commissioning of the Southern African Large Telescope}.
p.~0

\bibitem[\protect\citeauthoryear{{Capaccioli} \& {Schipani}}{{Capaccioli} \&
  {Schipani}}{2011}]{Capaccioli2011}
{Capaccioli} M.,  {Schipani} P.,  2011, The Messenger, 146, 2

\bibitem[\protect\citeauthoryear{{Carter} et~al.,}{{Carter}
  et~al.}{2013}]{Carter2013}
{Carter} P.~J.,  et~al., 2013, \mnras, 429, 2143

\bibitem[\protect\citeauthoryear{{Castelli} \& {Kurucz}}{{Castelli} \&
  {Kurucz}}{2004}]{Castelli2004}
{Castelli} F.,  {Kurucz} R.~L.,  2004, ArXiv Astrophysics e-prints

\bibitem[\protect\citeauthoryear{{Chang}, {Protopapas}, {Kim} \&
  {Byun}}{{Chang} et~al.}{2013}]{Chang2013}
{Chang} S.-W.,  {Protopapas} P.,  {Kim} D.-W.,    {Byun} Y.-I.,  2013, \aj,
  145, 132

\bibitem[\protect\citeauthoryear{{Crawford} et~al.,}{{Crawford}
  et~al.}{2010}]{Crawford2014}
{Crawford} S.~M.,  et~al., 2010, in Society of Photo-Optical Instrumentation
  Engineers (SPIE) Conference Series Vol.~7737 of Society of Photo-Optical
  Instrumentation Engineers (SPIE) Conference Series, {PySALT: the SALT science
  pipeline}.
p.~25

\bibitem[\protect\citeauthoryear{{Drake} et~al.,}{{Drake}
  et~al.}{2009}]{Drake2009}
{Drake} A.~J.,  et~al., 2009, \apj, 696, 870

\bibitem[\protect\citeauthoryear{{Drew} et~al.,}{{Drew}
  et~al.}{2014}]{Drew2014}
{Drew} J.~E.,  et~al., 2014, \mnras, 440, 2036

\bibitem[\protect\citeauthoryear{{Graham}, {Drake}, {Djorgovski}, {Mahabal},
  {Donalek}, {Duan} \& {Maker}}{{Graham} et~al.}{2013}]{Graham2013}
{Graham} M.~J.,  {Drake} A.~J.,  {Djorgovski} S.~G.,  {Mahabal} A.~A.,
  {Donalek} C.,  {Duan} V.,    {Maker} A.,  2013, \mnras, 434, 3423

\bibitem[\protect\citeauthoryear{{Groot} et~al.,}{{Groot}
  et~al.}{2003}]{Groot2003}
{Groot} P.~J.,  et~al., 2003, \mnras, 339, 427

\bibitem[\protect\citeauthoryear{{Hartman}, {Gaudi}, {Holman}, {McLeod},
  {Stanek}, {Barranco}, {Pinsonneault} \& {Kalirai}}{{Hartman}
  et~al.}{2008}]{Hartman2008}
{Hartman} J.~D.,  {Gaudi} B.~S.,  {Holman} M.~J.,  {McLeod} B.~A.,  {Stanek}
  K.~Z.,  {Barranco} J.~A.,  {Pinsonneault} M.~H.,    {Kalirai} J.~S.,  2008,
  \apj, 675, 1254

\bibitem[\protect\citeauthoryear{{Henden}, {Welch}, {Terrell} \&
  {Levine}}{{Henden} et~al.}{2009}]{Henden2009}
{Henden} A.~A.,  {Welch} D.~L.,  {Terrell} D.,    {Levine} S.~E.,  2009, in
  American Astronomical Society Meeting Abstracts 214 Vol.~214 of American
  Astronomical Society Meeting Abstracts, {The AAVSO Photometric All-Sky Survey
  (APASS)}.
p. 407.02

\bibitem[\protect\citeauthoryear{{Hermes} et~al.,}{{Hermes}
  et~al.}{2012}]{Hermes2012}
{Hermes} J.~J.,  et~al., 2012, \apjl, 757, L21

\bibitem[\protect\citeauthoryear{{Israel}, {Panzera}, {Campana}, {Lazzati},
  {Covino}, {Tagliaferri} \& {Stella}}{{Israel} et~al.}{1999}]{Israel1999}
{Israel} G.~L.,  {Panzera} M.~R.,  {Campana} S.,  {Lazzati} D.,  {Covino} S.,
  {Tagliaferri} G.,    {Stella} L.,  1999, \aap, 349, L1

\bibitem[\protect\citeauthoryear{{Jarrett}, {Chester}, {Cutri}, {Schneider},
  {Skrutskie} \& {Huchra}}{{Jarrett} et~al.}{2000}]{Jarrett2000}
{Jarrett} T.~H.,  {Chester} T.,  {Cutri} R.,  {Schneider} S.,  {Skrutskie} M.,
    {Huchra} J.~P.,  2000, \aj, 119, 2498

\bibitem[\protect\citeauthoryear{{Jonker} et~al.,}{{Jonker}
  et~al.}{2011}]{Jonker2011}
{Jonker} P.~G.,  et~al., 2011, \apjs, 194, 18

\bibitem[\protect\citeauthoryear{{Kilic}, {Brown}, {Gianninas}, {Hermes},
  {Allende Prieto} \& {Kenyon}}{{Kilic} et~al.}{2014}]{Kilic2014}
{Kilic} M.,  {Brown} W.~R.,  {Gianninas} A.,  {Hermes} J.~J.,  {Allende Prieto}
  C.,    {Kenyon} S.~J.,  2014, ArXiv e-prints

\bibitem[\protect\citeauthoryear{{Kilic}, {Brown}, {Hermes}, {Allende Prieto},
  {Kenyon}, {Winget} \& {Winget}}{{Kilic} et~al.}{2011}]{Kilic2011}
{Kilic} M.,  {Brown} W.~R.,  {Hermes} J.~J.,  {Allende Prieto} C.,  {Kenyon}
  S.~J.,  {Winget} D.~E.,    {Winget} K.~I.,  2011, \mnras, 418, L157

\bibitem[\protect\citeauthoryear{{Kim} et~al.,}{{Kim}  et~al.}{2010}]{Kim2010}
{Kim} D.-W.,  et~al., 2010, \aj, 139, 757

\bibitem[\protect\citeauthoryear{{Kobulnicky}, {Nordsieck}, {Burgh}, {Smith},
  {Percival}, {Williams} \& {O'Donoghue}}{{Kobulnicky}
  et~al.}{2003}]{Kobulnicky2003}
{Kobulnicky} H.~A.,  {Nordsieck} K.~H.,  {Burgh} E.~B.,  {Smith} M.~P.,
  {Percival} J.~W.,  {Williams} T.~B.,    {O'Donoghue} D.,  2003, in {Iye} M.,
  {Moorwood} A.~F.~M.,  eds, Instrument Design and Performance for
  Optical/Infrared Ground-based Telescopes Vol.~4841 of Society of
  Photo-Optical Instrumentation Engineers (SPIE) Conference Series, {Prime
  focus imaging spectrograph for the Southern African large telescope:
  operational modes}.
pp 1634--1644

\bibitem[\protect\citeauthoryear{{Kuijken}}{{Kuijken}}{2011}]{Kuijken2011}
{Kuijken} K.,  2011, The Messenger, 146, 8

\bibitem[\protect\citeauthoryear{{Lang}, {Hogg}, {Mierle}, {Blanton} \&
  {Roweis}}{{Lang} et~al.}{2010}]{Lang2010}
{Lang} D.,  {Hogg} D.~W.,  {Mierle} K.,  {Blanton} M.,    {Roweis} S.,  2010,
  \aj, 139, 1782

\bibitem[\protect\citeauthoryear{{Law} et~al.,}{{Law}  et~al.}{2009}]{Law2009}
{Law} N.~M.,  et~al., 2009, \pasp, 121, 1395

\bibitem[\protect\citeauthoryear{{Levitan} et~al.,}{{Levitan}
  et~al.}{2011}]{Levitan2011}
{Levitan} D.,  et~al., 2011, \apj, 739, 68

\bibitem[\protect\citeauthoryear{{Levitan}, {Groot}, {Prince}, {Kulkarni},
  {Laher}, {Ofek}, {Sesar} \& {Surace}}{{Levitan} et~al.}{2015}]{Levitan2015}
{Levitan} D.,  {Groot} P.~J.,  {Prince} T.~A.,  {Kulkarni} S.~R.,  {Laher} R.,
  {Ofek} E.~O.,  {Sesar} B.,    {Surace} J.,  2015, \mnras, 446, 391

\bibitem[\protect\citeauthoryear{{Lomb}}{{Lomb}}{1976}]{Lomb1976}
{Lomb} N.~R.,  1976, \apss, 39, 447

\bibitem[\protect\citeauthoryear{{McNamara}, {Clementini} \&
  {Marconi}}{{McNamara} et~al.}{2007}]{McNamara2007}
{McNamara} D.~H.,  {Clementini} G.,    {Marconi} M.,  2007, \aj, 133, 2752

\bibitem[\protect\citeauthoryear{{Morales-Rueda}, {Groot}, {Augusteijn},
  {Nelemans}, {Vreeswijk} \& {van den Besselaar}}{{Morales-Rueda}
  et~al.}{2006}]{Morales2006}
{Morales-Rueda} L.,  {Groot} P.~J.,  {Augusteijn} T.,  {Nelemans} G.,
  {Vreeswijk} P.~M.,    {van den Besselaar} E.~J.~M.,  2006, \mnras, 371, 1681

\bibitem[\protect\citeauthoryear{{Morales-Rueda} \& {Marsh}}{{Morales-Rueda} \&
  {Marsh}}{2002}]{Morales2002}
{Morales-Rueda} L.,  {Marsh} T.~R.,  2002, \mnras, 332, 814

\bibitem[\protect\citeauthoryear{{Nelemans}, {Portegies Zwart}, {Verbunt} \&
  {Yungelson}}{{Nelemans} et~al.}{2001}]{Nelemans2001b}
{Nelemans} G.,  {Portegies Zwart} S.~F.,  {Verbunt} F.,    {Yungelson} L.~R.,
  2001, \aap, 368, 939

\bibitem[\protect\citeauthoryear{{Petersen} \&
  {Christensen-Dalsgaard}}{{Petersen} \&
  {Christensen-Dalsgaard}}{1999}]{Peterson1999}
{Petersen} J.~O.,  {Christensen-Dalsgaard} J.,  1999, \aap, 352, 547

\bibitem[\protect\citeauthoryear{{Pollacco} et~al.,}{{Pollacco}
  et~al.}{2006}]{Pollacco2006}
{Pollacco} D.,  et~al., 2006, \apss, 304, 253

\bibitem[\protect\citeauthoryear{{Ramsay} et~al.,}{{Ramsay}
  et~al.}{2009}]{Ramsay2009}
{Ramsay} G.,  et~al., 2009, \mnras, 398, 1333

\bibitem[\protect\citeauthoryear{{Ramsay} et~al.,}{{Ramsay}
  et~al.}{2014}]{Ramsay2014}
{Ramsay} G.,  et~al., 2014, \mnras, 437, 132

\bibitem[\protect\citeauthoryear{{Ramsay} \& {Hakala}}{{Ramsay} \&
  {Hakala}}{2005}]{Ramsay2005}
{Ramsay} G.,  {Hakala} P.,  2005, \mnras, 360, 314

\bibitem[\protect\citeauthoryear{{Ramsay}, {Hakala} \& {Cropper}}{{Ramsay}
  et~al.}{2002}]{Ramsay2002}
{Ramsay} G.,  {Hakala} P.,    {Cropper} M.,  2002, \mnras, 332, L7

\bibitem[\protect\citeauthoryear{{Ramsay}, {Napiwotzki}, {Hakala} \&
  {Lehto}}{{Ramsay} et~al.}{2006}]{Ramsay2006}
{Ramsay} G.,  {Napiwotzki} R.,  {Hakala} P.,    {Lehto} H.,  2006, \mnras, 371,
  957

\bibitem[\protect\citeauthoryear{{Rappaport}, {Joss} \& {Webbink}}{{Rappaport}
  et~al.}{1982}]{Rappaport1982}
{Rappaport} S.,  {Joss} P.~C.,    {Webbink} R.~F.,  1982, \apj, 254, 616

\bibitem[\protect\citeauthoryear{{Rau}, {Roelofs}, {Groot}, {Marsh},
  {Nelemans}, {Steeghs}, {Salvato} \& {Kasliwal}}{{Rau} et~al.}{2010}]{Rau2010}
{Rau} A.,  {Roelofs} G.~H.~A.,  {Groot} P.~J.,  {Marsh} T.~R.,  {Nelemans} G.,
  {Steeghs} D.,  {Salvato} M.,    {Kasliwal} M.~M.,  2010, \apj, 708, 456

\bibitem[\protect\citeauthoryear{{Rodr{\'{\i}}guez} \&
  {L{\'o}pez-Gonz{\'a}lez}}{{Rodr{\'{\i}}guez} \&
  {L{\'o}pez-Gonz{\'a}lez}}{2000}]{Rodriguez2000}
{Rodr{\'{\i}}guez} E.,  {L{\'o}pez-Gonz{\'a}lez} M.~J.,  2000, \aap, 359, 597

\bibitem[\protect\citeauthoryear{{Roelofs} et~al.,}{{Roelofs}
  et~al.}{2009}]{Roelofs2009}
{Roelofs} G.~H.~A.,  et~al., 2009, \mnras, 394, 367

\bibitem[\protect\citeauthoryear{{Roelofs}, {Groot}, {Nelemans}, {Marsh} \&
  {Steeghs}}{{Roelofs} et~al.}{2006}]{Roelofs2006}
{Roelofs} G.~H.~A.,  {Groot} P.~J.,  {Nelemans} G.,  {Marsh} T.~R.,
  {Steeghs} D.,  2006, \mnras, 371, 1231

\bibitem[\protect\citeauthoryear{{Roelofs}, {Nelemans} \& {Groot}}{{Roelofs}
  et~al.}{2007}]{Roelofs2007}
{Roelofs} G.~H.~A.,  {Nelemans} G.,    {Groot} P.~J.,  2007, \mnras, 382, 685

\bibitem[\protect\citeauthoryear{{Scargle}}{{Scargle}}{1982}]{Scargle1982}
{Scargle} J.~D.,  1982, \apj, 263, 835

\bibitem[\protect\citeauthoryear{{Schwarzenberg-Czerny}}{{Schwarzenberg-Czerny%
}}{1989}]{SchwarzenbergCzerny1989}
{Schwarzenberg-Czerny} A.,  1989, \mnras, 241, 153

\bibitem[\protect\citeauthoryear{{Smak}}{{Smak}}{1967}]{Smak1967}
{Smak} J.,  1967, Information Bulletin on Variable Stars, 182, 1

\bibitem[\protect\citeauthoryear{{Solheim}}{{Solheim}}{2010}]{Solheim2010}
{Solheim} J.-E.,  2010, \pasp, 122, 1133

\bibitem[\protect\citeauthoryear{{Stetson}}{{Stetson}}{1996}]{Stetson1996}
{Stetson} P.~B.,  1996, \pasp, 108, 851

\bibitem[\protect\citeauthoryear{{Tamuz}, {Mazeh} \& {North}}{{Tamuz}
  et~al.}{2006}]{Tamuz2006}
{Tamuz} O.,  {Mazeh} T.,    {North} P.,  2006, \mnras, 367, 1521

\bibitem[\protect\citeauthoryear{{Tamuz}, {Mazeh} \& {Zucker}}{{Tamuz}
  et~al.}{2005}]{Tamuz2005}
{Tamuz} O.,  {Mazeh} T.,    {Zucker} S.,  2005, \mnras, 356, 1466

\bibitem[\protect\citeauthoryear{{Tody}}{{Tody}}{1986}]{Tody1986}
{Tody} D.,  1986, in {Crawford} D.~L.,  ed., Instrumentation in astronomy VI
  Vol.~627 of Society of Photo-Optical Instrumentation Engineers (SPIE)
  Conference Series, {The IRAF Data Reduction and Analysis System}.
p.~733

\bibitem[\protect\citeauthoryear{{Tody}}{{Tody}}{1993}]{Tody1993}
{Tody} D.,  1993, in {Hanisch} R.~J.,  {Brissenden} R.~J.~V.,   {Barnes} J.,
  eds, Astronomical Data Analysis Software and Systems II Vol.~52 of
  Astronomical Society of the Pacific Conference Series, {IRAF in the
  Nineties}.
p.~173

\bibitem[\protect\citeauthoryear{{Wagner} et~al.,}{{Wagner}
  et~al.}{2014}]{Wagner2014}
{Wagner} R.~M.,  et~al., 2014, The Astronomer's Telegram, 6669, 1

\bibitem[\protect\citeauthoryear{{Warner} \& {Robinson}}{{Warner} \&
  {Robinson}}{1972}]{Warner1972}
{Warner} B.,  {Robinson} E.~L.,  1972, \mnras, 159, 101

\bibitem[\protect\citeauthoryear{{Warner} \& {Woudt}}{{Warner} \&
  {Woudt}}{2002}]{Warner2002}
{Warner} B.,  {Woudt} P.~A.,  2002, \pasp, 114, 129

\bibitem[\protect\citeauthoryear{{Wozniak}}{{Wozniak}}{2000}]{Wozniak2000}
{Wozniak} P.~R.,  2000, \actaa, 50, 421

\end{thebibliography}

\appendix
\section{OmegaWhite Semester 88 field coordinates and variable candidates}

\begin{table*}
\centering
 \caption{We note the sky co-ordinates for the center of all semester 88 fields together with the number of light curves for sources which lie in the range 13.5 mag $ < g < $ 21.5 mag.}
\begin{tabular}{llll}
 \hline
 Field ID & RA         & DEC         & Number of \\
          & (J2000)    & (J2000)     & Light Curves \\
\hline
 1a       & 07:38:57.9 & --29:39:20.1 & 57002 \\
 1b       & 07:43:34.0 & --29:39:15.8 & 62116 \\
 2a       & 07:48:15.9 & --29:39:20.1 & 54741 \\
 2b       & 07:52:52.1 & --29:39:15.9 & 59465 \\
 3a       & 07:57:33.9 & --29:39:20.1 & 49846 \\
 3b       & 08:02:10.0 & --29:39:15.9 & 56663 \\
 4a       & 08:06:51.9 & --29:39:20.1 & 64465 \\
 4b       & 08:11:28.1 & --29:39:15.9 & 61218 \\
 5a       & 08:16:09.2 & --29:39:19.9 & 57718 \\
 5b       & 08:20:45.3 & --29:39:16.1 & 56695 \\
 6a       & 07:43:31.2 & --28:39:17.1 & 57812 \\
 6b       & 07:48:05.3 & --28:39:12.9 & 49519 \\
 8a       & 08:01:56.8 & --28:39:17.0 & 54573 \\
 8b       & 08:06:30.3 & --28:39:13.0 & 61541 \\
 9a       & 08:11:09.3 & --28:39:17.0 & 61184 \\
 9b       & 08:15:42.7 & --28:39:13.0 & 54986 \\
 11a      & 07:48:02.9 & --27:39:09.1 & 48046 \\
 11b      & 07:52:33.9 & --27:39:05.0 & 55484 \\
 12a      & 07:57:10.3 & --27:39:09.0 & 60343 \\
 12b      & 08:01:41.2 & --27:39:05.0 & 67633 \\
 13a      & 08:06:17.8 & --27:39:09.0 & 60572 \\
 13b      & 08:10:48.8 & --27:39:05.0 & 55318 \\
 14a      & 07:38:54.2 & --26:38:54.1 & 35784 \\
 14b      & 07:43:22.7 & --26:38:49.9 & 40929 \\
 15a      & 07:52:28.2 & --26:38:54.1 & 62493 \\
 15b      & 07:56:56.7 & --26:38:50.0 & 55150 \\
 \hline
 \end{tabular}
 \label{Appfields}
\end{table*}

\begin{table*}
\centering
\caption{Parameters of the 613 variable candidates (selected using $n$ = 15): star ID; RA and Dec; Lomb Scargle period (P$_{LS}$) and false alarm probability ($\log_{10}$(FAP)) of the highest peak in the periodogram, OW calibrated $g$-band magnitude, RMS of magnitude, VPHAS+ colour indices $g-r$ and $u-g$, and comments on variability type. }
\begin{tabular}[pos]{|l|c|c|r|c|c|c|r|r|c|c|}
\hline
 Star ID & RA (J2000) & DEC (J2000) & $P_{LS}$ & $\log_{10}$(FAP) & OW$g$ & RMS & $g-r$ & $u-g$  \\
 & (hh:mm:ss) & (\degr:\arcmin:\arcsec) &  (min) &  & (mag) & (mag) & (mag) & (mag) &  \\
\hline
OW\,J081124.34--295822.8 & 08:11:24.34 & --29:58:22.8 & 6.47 & --3.31 & 18.04 & 0.033 &   &   & \\
OW\,J080916.81--281226.2 & 08:09:16.81 & --28:12:26.2 & 6.75 & --3.04 & 18.13 & 0.008 & 0.83 & 0.55 & \\
OW\,J075449.27--263259.3 & 07:54:49.27 & --26:32:59.3 & 6.98 & --3.34 & 20.50 & 0.087 & 1.17 &   & \\
OW\,J075833.40--270202.5 & 07:58:33.40 & --27:02:02.5 & 7.46 & --3.21 & 20.61 & 0.110 & 1.13 &   & \\
OW\,J074809.65--291214.1 & 07:48:09.65 & --29:12:14.1 & 8.30 & --3.34 & 16.88 & 0.008 &  &  & \\
OW\,J080731.83--274915.1 & 08:07:31.83 & --27:49:15.1 & 8.34 & --3.29 & 17.08 & 0.004 & 0.82 & 0.60 & \\
\hline
\end{tabular}
\label{lc_parameters_full}
\end{table*}

\begin{table*}
\centering
\contcaption{Parameters of the 613 variable candidates (selected using $n$ = 15): star ID; RA and Dec; Lomb Scargle period (P$_{LS}$) and false alarm probability ($\log_{10}$(FAP)) of the highest peak in the periodogram, OW calibrated $g$-band magnitude, RMS of magnitude, VPHAS+ colour indices $g-r$ and $u-g$, and comments on variability type. }
\begin{tabular}[pos]{|l|c|c|r|c|c|c|r|r|c|c|}
\hline
 Star ID & RA (J2000) & DEC (J2000) & $P_{LS}$ & $\log_{10}$(FAP) & OW$g$ & RMS & $g-r$ & $u-g$  \\
 & (hh:mm:ss) & (\degr:\arcmin:\arcsec) &  (min) &  & (mag) & (mag) & (mag) & (mag) &  \\
\hline
OW\,J074908.65--275805.7 & 07:49:08.65 & --27:58:05.7 & 32.18 & --3.12 & 14.34 & 0.003 & 0.32 & 0.52 & \\
OW\,J080901.69--295231.7 & 08:09:01.69 & --29:52:31.7 & 32.61 & --4.08 & 14.59 & 0.005 & 0.26 & 0.52 & \\
OW\,J074233.32--261325.8 & 07:42:33.32 & --26:13:25.8 & 33.87 & --3.21 & 14.32 & 0.003 & 0.29 & 0.47 & \\
OW\,J080416.01--290712.7 & 08:04:16.01 & --29:07:12.7 & 34.09 & --3.02 & 16.54 & 0.006 &   &   & \\
OW\,J074626.36--274840.5 & 07:46:26.36 & --27:48:40.5 & 34.71 & --3.02 & 16.50 & 0.005 & 1.51 & 1.65 & \\
OW\,J081106.60--273648.4 & 08:11:06.60 & --27:36:48.4 & 35.48 & --3.72 & 15.34 & 0.004 &   &   & \\
\hline
\end{tabular}
\end{table*}

\begin{table*}
\centering
\contcaption{Parameters of the 613 variable candidates (selected using $n$ = 15): star ID; RA and Dec; Lomb Scargle period (P$_{LS}$) and false alarm probability ($\log_{10}$(FAP)) of the highest peak in the periodogram, OW calibrated $g$-band magnitude, RMS of magnitude, VPHAS+ colour indices $g-r$ and $u-g$, and comments on variability type. }
\begin{tabular}[pos]{|l|c|c|r|c|c|c|r|r|c|c|} 
\hline
Star ID & RA (J2000) & DEC (J2000) & $P_{LS}$ & $\log_{10}$(FAP) & OW$g$ & RMS & $g-r$ & $u-g$  \\
 & (hh:mm:ss) & (\degr:\arcmin:\arcsec) &  (min) &  & (mag) & (mag) & (mag) & (mag) &  \\
\hline
OW\,J073725.87--263356.4 & 07:37:25.87 & --26:33:56.4 & 42.92 & --3.59 & 15.16 & 0.004 & 0.50 & 0.57 & \\
OW\,J075313.21--273742.4 & 07:53:13.21 & --27:37:42.4 & 42.99 & --2.73 & 14.41 & 0.003 & 0.47 & 0.44 & \\
OW\,J080304.91--273919.9 & 08:03:04.91 & --27:39:19.9 & 43.13 & --4.31 & 17.61 & 0.012 & 0.44 & 0.58 & \\
OW\,J075804.98--293133.2 & 07:58:04.98 & --29:31:33.2 & 43.15 & --3.49 & 14.12 & 0.002 &  &  & \\
OW\,J074637.60--274453.0 & 07:46:37.60 & --27:44:53.0 & 43.23 & --2.59 & 17.16 & 0.011 & 0.71 & 0.79 & \\
OW\,J074537.63--263020.0 & 07:45:37.63 & --26:30:20.0 & 43.28 & --2.85 & 15.96 & 0.007 & 0.48 & 0.61 & \\
\hline
\end{tabular}
\end{table*}

\begin{table*}
\centering
\contcaption{Parameters of the 613 variable candidates (selected using $n$ = 15): star ID; RA and Dec; Lomb Scargle period (P$_{LS}$) and false alarm probability ($\log_{10}$(FAP)) of the highest peak in the periodogram, OW calibrated $g$-band magnitude, RMS of magnitude, VPHAS+ colour indices $g-r$ and $u-g$, and comments on variability type. }
\begin{tabular}[pos]{|l|c|c|r|c|c|c|r|r|c|c|}
\hline
Star ID & RA (J2000) & DEC (J2000) & $P_{LS}$ & $\log_{10}$(FAP) & OW$g$ & RMS & $g-r$ & $u-g$  \\
 & (hh:mm:ss) & (\degr:\arcmin:\arcsec) &  (min) &  & (mag) & (mag) & (mag) & (mag) &  \\
\hline
OW\,J073842.50--264752.1 & 07:38:42.50 & --26:47:52.1 & 51.47 & --3.98 & 16.01 & 0.018 & 0.62 & 0.75 & \\
OW\,J074017.09--294217.3 & 07:40:17.09 & --29:42:17.3 & 51.49 & --3.52 & 19.90 & 0.067 & 1.18 & 1.36 & \\
OW\,J074343.36--295605.4 & 07:43:43.36 & --29:56:05.4 & 51.53 & --4.74 & 15.79 & 0.007 & 0.64 & 0.80 & \\
OW\,J074209.66--294726.9 & 07:42:09.66 & --29:47:26.9 & 51.53 & --3.68 & 16.19 & 0.004 & 0.59 & 0.65 & \\
OW\,J074316.36--293113.8 & 07:43:16.36 & --29:31:13.8 & 51.71 & --3.58 & 18.67 & 0.031 & 0.95 & 0.74 & \\
OW\,J080550.52--275257.5 & 08:05:50.52 & --27:52:57.5 & 51.88 & --3.10 & 17.14 & 0.008 & 0.62 & 0.59 & \\
\hline
\end{tabular}
\end{table*}

\begin{table*}
\centering
\contcaption{Parameters of the 613 variable candidates (selected using $n$ = 15): star ID; RA and Dec; Lomb Scargle period (P$_{LS}$) and false alarm probability ($\log_{10}$(FAP)) of the highest peak in the periodogram, OW calibrated $g$-band magnitude, RMS of magnitude, VPHAS+ colour indices $g-r$ and $u-g$, and comments on variability type. }
\begin{tabular}[pos]{|l|c|c|r|c|c|c|r|r|c|c|}
\hline
Star ID & RA (J2000) & DEC (J2000) & $P_{LS}$ & $\log_{10}$(FAP) & OW$g$ & RMS & $g-r$ & $u-g$  \\
 & (hh:mm:ss) & (\degr:\arcmin:\arcsec) &  (min) &  & (mag) & (mag) & (mag) & (mag) &  \\
\hline
OW\,J075009.03--292023.1 & 07:50:09.03 & --29:20:23.1 & 59.21 & --3.14 & 17.39 & 0.009 &   &   & \\
OW\,J074924.79--280357.8 & 07:49:24.79 & --28:03:57.8 & 59.40 & --2.98 & 17.01 & 0.009 & 0.87 & 0.96 & \\
OW\,J080008.72--293132.7 & 08:00:08.72 & --29:31:32.7 & 59.72 & --4.36 & 19.19 & 0.087 &  &  & \\
OW\,J074029.02--300434.9 & 07:40:29.02 & --30:04:34.9 & 60.07 & --3.96 & 19.48 & 0.381 &   & 0.68 & \\
OW\,J080310.43--290316.8 & 08:03:10.43 & --29:03:16.8 & 60.08 & --3.43 & 13.76 & 0.011 &   &   & \\
OW\,J081102.39--291220.8 & 08:11:02.39 & --29:12:20.8 & 60.15 & --4.18 & 16.60 & 0.012 &   &   & \\
\hline
\end{tabular}
\end{table*}

\begin{table*}
\centering
\contcaption{Parameters of the 613 variable candidates (selected using $n$ = 15): star ID; RA and Dec; Lomb Scargle period (P$_{LS}$) and false alarm probability ($\log_{10}$(FAP)) of the highest peak in the periodogram, OW calibrated $g$-band magnitude, RMS of magnitude, VPHAS+ colour indices $g-r$ and $u-g$, and comments on variability type. }
\begin{tabular}[pos]{|l|c|c|r|c|c|c|r|r|c|c|}  
\hline
Star ID & RA (J2000) & DEC (J2000) & $P_{LS}$ & $\log_{10}$(FAP) & OW$g$ & RMS & $g-r$ & $u-g$  \\
 & (hh:mm:ss) & (\degr:\arcmin:\arcsec) &  (min) &  & (mag) & (mag) & (mag) & (mag) &  \\
\hline
OW\,J075026.49--273513.6 & 07:50:26.49 & --27:35:13.6 & 65.54 & --4.43 & 15.20 & 0.004 & 0.50 & 0.52 & \\
OW\,J080625.13--293011.0 & 08:06:25.13 & --29:30:11.0 & 65.69 & --4.71 & 15.90 & 0.008 & 0.53 & 0.45 & \\
OW\,J074033.58--263001.7 & 07:40:33.58 & --26:30:01.7 & 65.75 & --3.93 & 14.67 & 0.009 &   &   & \\
OW\,J074517.66--264834.6 & 07:45:17.66 & --26:48:34.6 & 65.79 & --3.07 & 15.91 & 0.005 & 0.56 & 0.48 & \\
OW\,J074352.62--300512.2 & 07:43:52.62 & --30:05:12.2 & 65.79 & --4.46 & 15.70 & 0.005 & 0.61 & 0.66 & \\
OW\,J073843.80--293811.7 & 07:38:43.80 & --29:38:11.7 & 65.82 & --4.57 & 16.24 & 0.004 & 0.80 & 0.77 & \\
\hline
\end{tabular}
\end{table*}

\begin{table*}
\centering
\contcaption{Parameters of the 613 variable candidates (selected using $n$ = 15): star ID; RA and Dec; Lomb Scargle period (P$_{LS}$) and false alarm probability ($\log_{10}$(FAP)) of the highest peak in the periodogram, OW calibrated $g$-band magnitude, RMS of magnitude, VPHAS+ colour indices $g-r$ and $u-g$, and comments on variability type. }
\begin{tabular}[pos]{|l|c|c|r|c|c|c|r|r|c|c|}
\hline
Star ID & RA (J2000) & DEC (J2000) & $P_{LS}$ & $\log_{10}$(FAP) & OW$g$ & RMS & $g-r$ & $u-g$  \\
 & (hh:mm:ss) & (\degr:\arcmin:\arcsec) &  (min) &  & (mag) & (mag) & (mag) & (mag) &  \\
\hline
OW\,J073929.00--295950.4 & 07:39:29.00 & --29:59:50.4 & 71.40 & --3.63 & 16.32 & 0.008 & 0.70 & 0.74 & \\
OW\,J081119.48--280105.0 & 08:11:19.48 & --28:01:05.0 & 71.42 & --4.54 & 17.36 & 0.019 &   &   & \\
OW\,J080835.00--272656.8 & 08:08:35.00 & --27:26:56.8 & 71.89 & --3.81 & 16.23 & 0.020 & 0.49 & 0.63 & \\
OW\,J075907.07--263950.5 & 07:59:07.07 & --26:39:50.5 & 72.23 & --3.96 & 18.83 & 0.053 & 0.85 & 0.48 & \\
OW\,J075449.09--261721.2 & 07:54:49.09 & --26:17:21.2 & 72.23 & --3.65 & 17.42 & 0.010 &   &   & \\
OW\,J080836.11--290207.3 & 08:08:36.11 & --29:02:07.3 & 72.63 & --3.02 & 16.47 & 0.006 &   &   & \\
\hline
\end{tabular}
\end{table*}

\begin{table*}
\centering
\contcaption{Parameters of the 613 variable candidates (selected using $n$ = 15): star ID; RA and Dec; Lomb Scargle period (P$_{LS}$) and false alarm probability ($\log_{10}$(FAP)) of the highest peak in the periodogram, OW calibrated $g$-band magnitude, RMS of magnitude, VPHAS+ colour indices $g-r$ and $u-g$, and comments on variability type. }
\begin{tabular}[pos]{|l|c|c|r|c|c|c|r|r|c|c|}
\hline
Star ID & RA (J2000) & DEC (J2000) & $P_{LS}$ & $\log_{10}$(FAP) & OW$g$ & RMS & $g-r$ & $u-g$  \\
 & (hh:mm:ss) & (\degr:\arcmin:\arcsec) &  (min) &  & (mag) & (mag) & (mag) & (mag) &  \\
\hline
OW\,J081258.21--300829.3 & 08:12:58.21 & --30:08:29.3 & 79.75 & --4.66 & 15.68 & 0.020 &   &   & \\
OW\,J081243.44--274754.9 & 08:12:43.44 & --27:47:54.9 & 80.28 & --4.33 & 17.10 & 0.009 &   &   & \\
OW\,J080909.14--273354.5 & 08:09:09.14 & --27:33:54.5 & 80.28 & --3.68 & 17.05 & 0.010 & 0.60 & 0.58 & \\
OW\,J081334.04--300054.5 & 08:13:34.04 & --30:00:54.5 & 80.48 & --4.36 & 19.67 & 0.160 &   &   & \\
OW\,J074317.66--300712.2 & 07:43:17.66 & --30:07:12.2 & 80.53 & --6.60 & 15.03 & 0.006 & 0.54 & 0.85 & \\
OW\,J081557.95--293824.6 & 08:15:57.95 & --29:38:24.6 & 80.66 & --5.27 & 15.04 & 0.007 &   &   & \\
\hline
\end{tabular}
\end{table*}

\begin{table*}
\centering
\contcaption{Parameters of the 613 variable candidates (selected using $n$ = 15): star ID; RA and Dec; Lomb Scargle period (P$_{LS}$) and false alarm probability ($\log_{10}$(FAP)) of the highest peak in the periodogram, OW calibrated $g$-band magnitude, RMS of magnitude, VPHAS+ colour indices $g-r$ and $u-g$, and comments on variability type. }
\begin{tabular}[pos]{|l|c|c|r|c|c|c|r|r|c|c|}
\hline
Star ID & RA (J2000) & DEC (J2000) & $P_{LS}$ & $\log_{10}$(FAP) & OW$g$ & RMS & $g-r$ & $u-g$  \\
 & (hh:mm:ss) & (\degr:\arcmin:\arcsec) &  (min) &  & (mag) & (mag) & (mag) & (mag) &  \\
\hline
OW\,J073909.96--300620.0 & 07:39:09.96 & --30:06:20.0 & 89.57 & --5.92 & 13.64 & 0.004 & 0.48 & 0.55 &  \\
OW\,J075007.25--290929.5 & 07:50:07.25 & --29:09:29.5 & 89.61 & --4.09 & 18.30 & 0.028 &  &  &  \\
OW\,J074438.06--293024.4 & 07:44:38.06 & --29:30:24.4 & 89.87 & --5.37 & 16.23 & 0.015 &  &  &  \\
OW\,J080951.45--280013.6 & 08:09:51.45 & --28:00:13.6 & 90.16 & --3.36 & 17.08 & 0.013 & 0.67 & 0.66 &  \\
OW\,J081234.75--280445.8 & 08:12:34.75 & --28:04:45.8 & 90.16 & --3.88 & 20.61 & 0.190 &   &   &  \\
OW\,J075142.86--274827.4 & 07:51:42.86 & --27:48:27.4 & 90.22 & --3.71 & 14.79 & 0.003 &   &   &  \\
\hline
\end{tabular}
\end{table*}

\begin{table*}
\centering
\contcaption{Parameters of the 613 variable candidates (selected using $n$ = 15): star ID; RA and Dec; Lomb Scargle period (P$_{LS}$) and false alarm probability ($\log_{10}$(FAP)) of the highest peak in the periodogram, OW calibrated $g$-band magnitude, RMS of magnitude, VPHAS+ colour indices $g-r$ and $u-g$, and comments on variability type. }
\begin{tabular}[pos]{|l|c|c|r|c|c|c|r|r|c|c|}
\hline
Star ID & RA (J2000) & DEC (J2000) & $P_{LS}$ & $\log_{10}$(FAP) & OW$g$ & RMS & $g-r$ & $u-g$  \\
 & (hh:mm:ss) & (\degr:\arcmin:\arcsec) &  (min) &  & (mag) & (mag) & (mag) & (mag) &  \\
\hline
OW\,J075221.80--293027.7 & 07:52:21.80 & --29:30:27.7 & 99.14 & --3.41 & 15.65 & 0.005 &  &  & \\
OW\,J075031.81--294516.7 & 07:50:31.81 & --29:45:16.7 & 99.15 & --3.53 & 14.84 & 0.006 &  &  & \\
OW\,J074606.57--295609.2 & 07:46:06.57 & --29:56:09.2 & 99.74 & --3.50 & 15.71 & 0.003 & 0.49 & 0.65 & \\
OW\,J075011.62--291209.3 & 07:50:11.62 & --29:12:09.3 & 99.74 & --4.44 & 17.04 & 0.017 &  &  & \\
OW\,J075231.48--292530.5 & 07:52:31.48 & --29:25:30.5 & 99.75 & --3.54 & 17.67 & 0.015 &   &   & \\
OW\,J080620.34--275230.0 & 08:06:20.34 & --27:52:30.0 & 99.99 & --4.01 & 17.32 & 0.008 & 0.63 & 0.55 & \\
\hline
\end{tabular}
\end{table*}

\begin{table*}
\centering
\contcaption{Parameters of the 613 variable candidates (selected using $n$ = 15): star ID; RA and Dec; Lomb Scargle period (P$_{LS}$) and false alarm probability ($\log_{10}$(FAP)) of the highest peak in the periodogram, OW calibrated $g$-band magnitude, RMS of magnitude, VPHAS+ colour indices $g-r$ and $u-g$, and comments on variability type. }
\begin{tabular}[pos]{|l|c|c|r|c|c|c|r|r|c|c|} 
\hline
Star ID & RA (J2000) & DEC (J2000) & $P_{LS}$ & $\log_{10}$(FAP) & OW$g$ & RMS & $g-r$ & $u-g$  \\
 & (hh:mm:ss) & (\degr:\arcmin:\arcsec) &  (min) &  & (mag) & (mag) & (mag) & (mag) &  \\
\hline
OW\,J075802.18--263737.7 & 07:58:02.18 & --26:37:37.7 & 112.20 & --5.49 & 17.73 & 0.107 & 0.79 & 0.70 & \\
OW\,J080951.01--275210.3 & 08:09:51.01 & --27:52:10.3 & 112.24 & --4.14 & 19.98 & 0.140 & 1.00 & 0.64 & \\
OW\,J074714.30--291929.2 & 07:47:14.30 & --29:19:29.2 & 112.46 & --4.65 & 17.29 & 0.013 &  &  & \\
OW\,J074745.86--271601.8 & 07:47:45.86 & --27:16:01.8 & 113.27 & --5.38 & 17.77 & 0.053 & 1.06 & 1.06 & \\
OW\,J081143.07--275147.1 & 08:11:43.07 & --27:51:47.1 & 113.39 & --4.00 & 16.24 & 0.014 &   &   & \\
OW\,J081247.13--272220.9 & 08:12:47.13 & --27:22:20.9 & 113.39 & --5.64 & 19.15 & 0.100 &   &   & \\
\hline
\end{tabular}
\end{table*}

\label{lastpage}

\end{document}